\documentclass[twocolumn,showpacs,preprintnumbers,amsmath,amssymb,superscriptaddress,10pt,aps,prb]{revtex4-2}
\usepackage{epsfig,amsopn}
\usepackage{graphicx}
\usepackage{epstopdf}

\usepackage{float}
\usepackage{amsmath,amssymb}
\usepackage{amsthm}
\usepackage{tcolorbox}
\usepackage{enumerate}
\usepackage{xcolor}
\usepackage[normalem]{ulem}
\usepackage[colorlinks=true,linktoc=page,linkcolor=magenta,citecolor=magenta]{hyperref}

\DeclareMathOperator{\sgn}{sgn}

\newcommand\bea{\begin{eqnarray}}
	\newcommand\eea{\end{eqnarray}}
\newcommand\beq{\begin{equation}}  
	\newcommand\eeq{\end{equation}}

\begin{document}
\title{RKKY interaction in Weyl semimetal nanowires}
\author{Rohit Mukherjee}
\email{mukherjeemimo725@gmail.com}
\affiliation{Department of Physics, Indian Institute of Technology - Kanpur, Kanpur 208 016, India.}
\author{Asutosh Dubey}
\email{ashutosh@labs.iisertirupati.ac.in}
\affiliation{Department of Physics, Indian Institute of Technology - Kanpur, Kanpur 208 016, India.}
\affiliation{Department of Physics, Indian Institute of Science Education and Research Tirupati, Yerpedu Mandal 517 619, India. }

\begin{abstract}
We investigate the effective couplings induced between localized impurities on the surface of a Weyl semimetal (WSM) nanowire within the framework of Ruderman–Kittel–Kasuya–Yosida (RKKY) theory. The itinerant electrons from the chiral Fermi arc surface states mediate impurity-impurity interaction at low energies. As a result, the spin–momentum locking naturally plays a central role in shaping the spin–spin correlations. We show that the dominant interaction channels have distinct origins: while the azimuthal coupling, $J_{\phi\phi}$ term arises exclusively from Fermi arc states with identical spin polarization, the couplings $J_{\mu\nu}$ ($\mu,\nu = z,r$) are governed by Fermi arc states with opposite spin polarizations. Furthermore, we demonstrate that purely surface-mediated contributions exhibit different scaling behavior compared to those involving Fermi arcs and low-energy bulk states. We systematically untangle the contributions from bulk and surface states to the RKKY couplings, using analytical and numerical methods. Our results establish WSM nanowires as a versatile platform for engineering and simulating a broad class of spin models.

\end{abstract}

\maketitle

\section{Introduction}
\begin{figure}[tbph]
	\centering
	\includegraphics[width=0.6\linewidth]{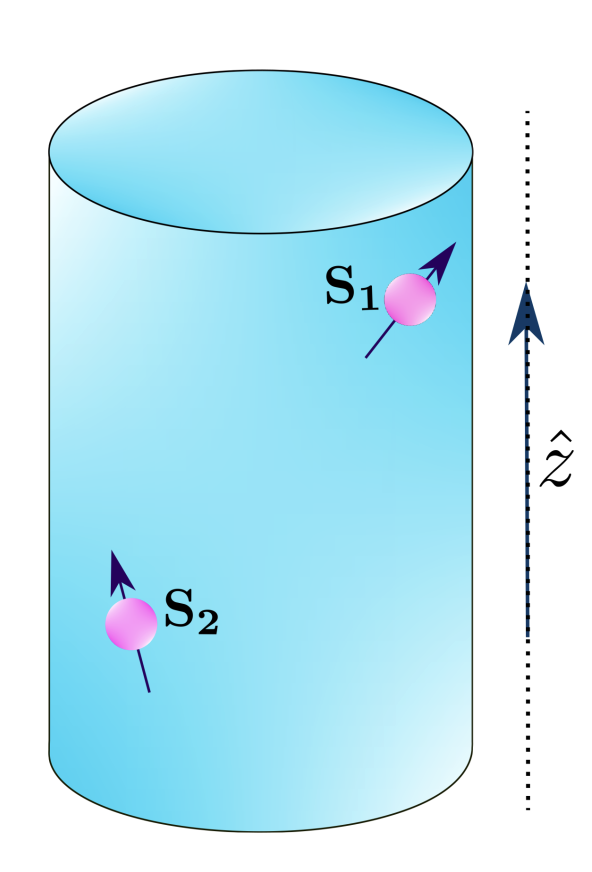}
	\caption{Schematic setup for the Weyl-semimetal cylindrical nanowire of radius $R$ and infinite length along the $\hat{z}$ direction. Impurity spins $\bold{S_{1}}$ and $\bold{S_{2}}$ are placed on the surface of the cylinder, and they can interact via the conduction electrons in the Weyl semimetal.}
	\label{fig:setupp}
\end{figure}

Weyl semimetals (WSMs) describe an extensive variety of three-dimensional topological quantum materials that host a gapless bulk spectrum around a set of non-trivial points in the Brillouin zone, known as the \textit{Weyl nodes}~\cite{armitage2018weyl,jia2016weyl,rao2016weyl,PhysRevLett.107.127205}. The low-energy spectrum around such nodes follows the equation of motion characteristic of $Weyl fermions$. The Weyl nodes of opposing chirality are known to be separated in momentum space, and in a finite geometry they are connected by an exotic non-closed surface state, known as the \textit{Fermi arcs}~\cite{arca,arcb,arcc,arcd}. This pair of Weyl nodes (with opposing chirality) acts as monopoles in the momentum space, behaving as a source or sink of the Berry curvature~\cite{nielsen1981absence1,nielsen1981absence2}. The low-energy excitations in Weyl semimetals (WSMs) are of particular interest because of their helical nature. The spin of itinerant electrons on the Fermi arc is locked to the momentum direction along the arc. This locking is the consequence of the chiral nature of Weyl fermions. Because of spin-momentum locking, the two electrons moving in the opposite direction along the Fermi arc must have opposite spin; as a result of this, it suppresses the backscattering of particular spin and hence leads to strong spin-dependent quantum correlations.
Numerous materials are predicted to be suitable candidates for WSMs, and various experiments demonstrate their unique characteristics~\cite{lv2015experimental,xu2015experimental,deng2016experimental,PhysRevX.5.031023,lu2024}. The presence of spin-dependent quantum correlations influences many properties of these systems. In our work, we specifically focus on the implications of the effective spin correlations developed between magnetic impurities placed in the nanowire geometry of a WSM. %We resort to the Ruderman-Kittel-Kasuya-Yosida (RKKY) theory to study such phenomena. 

The RKKY theory~\cite{ruderman1954indirect,kasuya1956theory,yosida1957magnetic}  has long been a go-to method for studying such mechanisms developed between two localized impurity spins mediated by itinerant fermions, which is the WSM conduction electrons in our case. In spintronics materials, similar processes have gained a great deal of interest in recent years~\cite{spintronic1,spintronic2,spintronic3}.
The long-range spin-spin coupling is essential for the magnetic ordering of impurities and may also help in the understanding of the magnetic properties of the underlying system~\cite{hermenau2019stabilizing}. The RKKY interaction is determined by mobile carrier characteristics (e.g., chirality, energy dispersion, spinor structure, etc.). The magnitude of RKKY coupling oscillates with the distance ($D$) between impurities and decays as $1/D^\eta$, where $\eta$ relies on the density and effective mass of the charge carriers as well as other material-dependent characteristics. Within the systems of the solid-state materials, systems with large spin-orbit coupling~\cite{imamura2004twisted,schulz2009low}, particularly, topological and Dirac systems are interesting due to their ability to mediate long-range
interactions between spins~\cite{gao2009plane,liu2009magnetic,sherafati2011rkky,kogan2011rkky,mastrogiuseppe2016hybridization,zare2016topological,yang2016long,hsu2017nuclear,hsu2018effects,lee2015electrical,reja2017surface,reja2019spin,verma2019nonlocal}. Previous studies of RKKY interactions through bulk Weyl fermions demonstrated that interactions are not always isotropic and can be quite weak, but always exhibit hallmarks of the chiral nodes~\cite{chang2015rkky,sun2017rkky,hosseini2015ruderman}.
Recent studies have focused on RKKY interactions at the surfaces of Weyl semimetals (WSMs) in slab geometries, both semi-infinite and finite configurations~\cite{kasuya1956theory,duan2018indirect,ma2018kondo,verma2020rkky}. Work on carbon nanotubes and graphene nanoribbons has shown that tuning the Fermi level near the spin–orbit–induced gap can equalize different RKKY terms~\cite{graphenerkky1,graphenerkky2}. Meanwhile, transport in WSM nanowires has been studied~\cite{igarashi2017magnetotransport,PhysRevB.97.035429,kaladzhyan2019quantized,PhysRevB.104.155425}, revealing that under magnetic fields magnetotransport reflects the interplay between Fermi arc surface states and bulk Landau levels~\cite{igarashi2017magnetotransport}.

In this work, we investigate the effect of both Fermi arc surface states and low energy bulk states on the effective RKKY interaction between magnetic impurities localized on the surface of an infinitely long WSM nanowire. This device geometry is not only experimentally realizable but also analytically tractable, allowing for a comprehensive analysis of the interplay between Fermi arc states, bulk states and finite-size effects.
We explicitly calculate the contribution from the Fermi-arc surface states and the low energy bulk states and corroborate them with numerical analysis. Our study suggest that the symmetry of various RKKY correlation for surface states as well as low energy bulk states strongly depends upon the nature of Fermi arc and the geometry. Here, the sign and magnitude of different correlations depend on energy scales such as the Fermi energy and length scale such as separation between Fermi arcs. We also find that, for small radius, the bulk contribution to the coupling decays more slowly with increasing separation in the axial (\(\delta z\)) direction between impurity spins compared to the flatband Fermi-arc surface states.
Our effective RKKY exchange Hamiltonian, expressed in cylindrical coordinates, includes an anisotropic Heisenberg interaction, a Dzyaloshinskii–Moriya (DM) interaction, and a symmetric off-diagonal exchange term.

The rest of the paper is organized as follows, in Sec.~\ref{Formalism}, we introduce the model Hamiltonian of interest and calculate the analytical solutions for the wave functions in nanowire geometry. In Sec.~\ref{rkkyanalytic} we shortly discuss the basics of RKKY interaction and the technical tools that will be used to analyze it. Our numerical results are presented in Sec.~\ref{Sec:numericalresults} where we compare them with analytical results. Finally, we conclude with a summary and discussion in Sec.~\ref{discussion}.
 
\section{Model : Weyl Semimetal Nanowire}\label{Formalism}

\begin{figure}[tbph]
	\centering
	\includegraphics[width=1.0\linewidth]{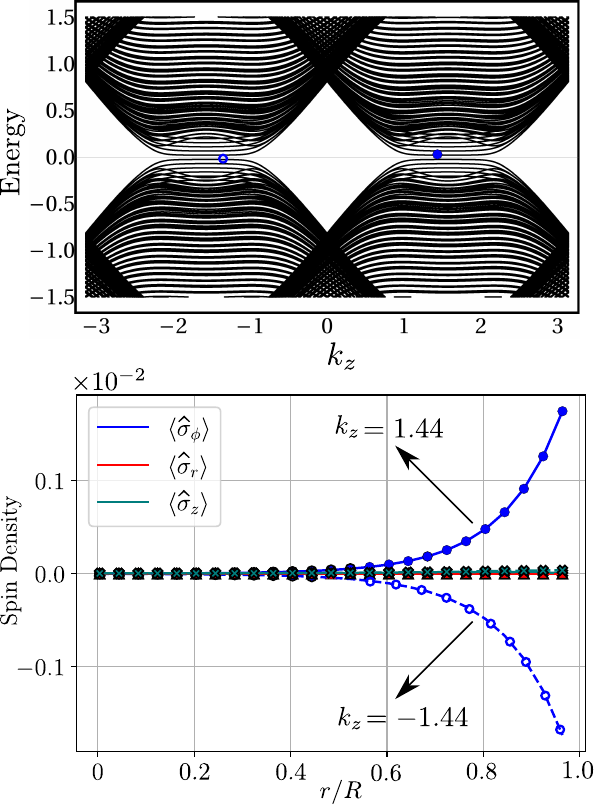}
	\caption{Top: WSM dispersion in nanowire geometry as a function of $k_{z}$. There are four Weyl nodes (located at $k_{z}=\pm k_{0}\pm \pi/2$, where $2k_{0}$ represents the length of each Fermi arc and $\pi$ is the separation between the center of the two Fermi arcs in momentum space) in bulk and two Fermi-arcs joining them. The two blue dots corresponds to the location of Weyl nodes at $\pm k_{z}$.The gap in the lowest energy band is proportional to the inverse of the radius $R$ of the nanowire. Bottom: Spin densities of the Fermi-arc surface states are shown as a function of radial distance $r$ for two values of $k_{z}$. The FA surface states are polarized along the $\hat{\phi}$ direction with $\langle \hat{\sigma}_{z}\rangle =\langle \hat{\sigma}_{r}\rangle=0$ and Fermi arcs with $k_{z}=\pm 1.44$ have exactly opposite spin polarization. The localization of the surface states depends primarily on the radius of the cylinder. All the lengths are measured in the unit of the lattice constant $a$, for our convenience we set $a=1$. Other parameters for the plot are as follows: $R$=25.0, $k_{0}=0.2\pi$,\ $v_{F}=1$.} 
	\label{fig:wsmnewdispersion}
\end{figure}
We start our discussion with a model Hamiltonian for the Weyl semimetal in cylindrical geometry as shown schematically in Fig.~(\ref{fig:setupp}). We assume that our system has a finite radius $R$ in the radial direction but is infinite along the $z$-axis. A minimal model of WSM breaks the time-reversal (TR) symmetry by having two Weyl nodes at the Fermi energy and we can use a two-band model to construct the low-energy effective Hamiltonian. If the two bands represent spin states, then the surface states (Fermi arcs) are spin-polarized, resulting in a fully spin-polarized nanowire.
%A two-band model can be used to construct the low-energy effective Hamiltonian of this model. If the two bands represent spin states, then the surface states (Fermi arcs) are spin-polarized, resulting in a fully spin-polarized nanowire.
However, the indirect spin-spin couplings, such as the RKKY coupling, are only of interest when the ground state is spin-unpolarized, resulting in dominating indirect exchange interaction~\cite{verma2020rkky}. As a result, our minimal model must include four Weyl nodes and two Fermi-arcs with opposite spin chirality. The low-energy effective Hamiltonian for this particular model in each chirality sector can be constructed as, 
\begin{equation}\label{eq:Hamiltonian}
H_{\xi}=v_{F} \big (\hat{p}_{x} \sigma_{x}  +\hat{p}_{y} \sigma_{y}\big) +M_{\xi}(k_{z})\sigma_{z},
\end{equation}
where $\xi=\pm 1$, $v_{F}$ is the Fermi velocity, $\hat{p}_{x}$ and $\hat{p}_{y}$ stands for the $x$ and $y$ component of momentum operator, $\sigma_{\nu}$ is the $\nu^{th}$ Pauli matrix.  We model the mass term $M_{\xi}$ in such a way that it breaks the TR symmetry in each chirality sector,
\begin{equation}
M_{\xi}(k_{z}) =
\begin{cases}
\xi(\cos{k_{0}}-\cos(k_{z}-\xi \pi/2))\kappa & \text{for $r<R$}\\
M_{0} & \text{for $r>R$},\\

\end{cases}       
\end{equation}
where $2k_{0}$ represents the length of each Fermi arc and $\pi$ is the separation between the center of the two Fermi arcs in momentum space, also we take $v_{F}=1$ and $\kappa = 1$ for the rest of the manuscript. In the above equation $r$ is the radial distance measured from the central axis of the nanowire. Although the TRS is broken for each block with $\xi=\pm 1$, two blocks are time-reversal partners of one another; hence the system as a whole maintains the TR symmetry. To compute the eigenvalues and eigenvectors of the Hamiltonian we write the Schr\"odinger equation for $r<R$ in cylindrical coordinate (the $\xi$ index is suppressed from this point onward for the sake of notational simplicity),
\begin{align}
\begin{bmatrix}
M(k_z) &- e^{-i\phi} \left(i\partial_r + \frac{1}{r} \partial_{\phi}\right) \\
e^{i\phi} \left(i\partial_r - \frac{1}{r} \partial_{\phi}\right) & -M(k_z)
\end{bmatrix} \Psi_{k_z}(r,\phi)\nonumber \\
=\epsilon({k_z}) \Psi_{k_z}(r,\phi),
\end{align}
where $\epsilon(k_{z})$ and $\Psi_{k_{z}}$ are the corresponding eigenenergy and eigenstate respectively. $\phi$ is the azimuthal coordinate of the nanowire.
The Hamiltonian in Eq.~(\ref{eq:Hamiltonian}) commutes with the current operator:
\begin{equation}
\begin{aligned}
\hat{j} &=-i\partial_{\phi}+\dfrac{1}{2}\sigma_{z},    
\end{aligned}
\end{equation}
as a result, operator $\hat{j}$ and the Hamiltonian in Eq.(\ref{eq:Hamiltonian}) have simultaneous eigenfunctions. We can choose them to have the following form,

\begin{equation}\label{wavefunctionwsm}
\begin{aligned}
\Psi_{k_{z}}(r,\phi) &=\sum_{m \in \mathbb{Z}} e^{im\phi} \begin{bmatrix}
u_{m}(k_{z},r) \\
e^{i\phi} v_{m}(k_{z},r)
\end{bmatrix}.    
\end{aligned}
\end{equation}

As we have a translational symmetry along the $\hat{z}$ direction, the $k_{z}$ is a good quantum number.
For $r<R$ we can write,
\begin{equation*}
\begin{aligned}
&\Big(\partial_{r}+\dfrac{m+1}{r} \Big) v_{m}^{<}=i \big(\epsilon(k_{z})-M(k_{z}) \big) u_{m}^{<},\\
&\Big(\partial_{r}-\dfrac{m}{r} \Big) u_{m}^{<}=i \big(\epsilon(k_{z})+M(k_{z}) \big) v_{m}^{<},    
\end{aligned}
\end{equation*}
whereas for $r>R$ one can write,
\begin{equation*}
\begin{aligned}
&\Big(\partial_{r}+\dfrac{m+1}{r} \Big) v_{m}^{>}=i \big(\epsilon(k_{z})-M_{0}) \big) u_{m}^{>},\\
&\Big(\partial_{r}-\dfrac{m}{r} \Big) u_{m}^{>}=i \big(\epsilon(k_{z})+M_{0}) \big) v_{m}^{>}.    
\end{aligned}
\end{equation*}
For $r<R$ appropriate solutions to these coupled equations have the form:
\begin{equation*}
u_{m}^{<}=J_{m}(Qr); \ \ \ \ v_{m}^{<}=\dfrac{iQ}{\epsilon(k_{z})+M(k_{z})} J_{m+1}(Qr),
\end{equation*}

where $J_{m}(.)$ is the Bessel function (first kind) of order $m$ and $Q=\sqrt{\epsilon(k_{z})^{2}-M^{2}(k_{z}) }$. For $r>R$, we have,

\begin{equation*}
u_{m}^{>}=H_{m}(Q_{0}R); \ \ \ \ v_{m}^{>}=\dfrac{iQ_{0}}{\epsilon(k_{z})+M_{0}} H_{m+1}(Q_{0}r),
\end{equation*}

with $Q_{0}=\sqrt{\epsilon({k_{z}})^{2}-M^{2}(k_{z})}$ $\to$ $iM_{0}$ as $M_{0} \to \infty$, $H_{m}(\cdot)$ is the Hankel function (first kind) of order $m$. Now matching the wavefunctions at the boundary of the cylinder provides us the solution of the dispersion $\epsilon({k_{z}})$,

\begin{equation}\label{transdental1}
\dfrac{\epsilon({k_{z}})+M(k_{z})}{Q}\dfrac{J_{m}(QR)}{J_{m+1}(QR)}=\sgn(M_{0}).
\end{equation} 

Now, we restore the chirality index $\xi$, the transcendental equation (\ref{transdental1}) takes the form,

\begin{equation}\label{transdental2}
\dfrac{\epsilon_{\xi}(k_{z})+M_{\xi}(k_{z})}{Q_{\xi}}\dfrac{J_{m}(Q_{\xi}R)}{J_{m+1}(Q_{\xi}R)}=\sgn(M_{0}).
\end{equation} 

\begin{figure*}[t]
	\centering
	\includegraphics[width=1.0\linewidth]{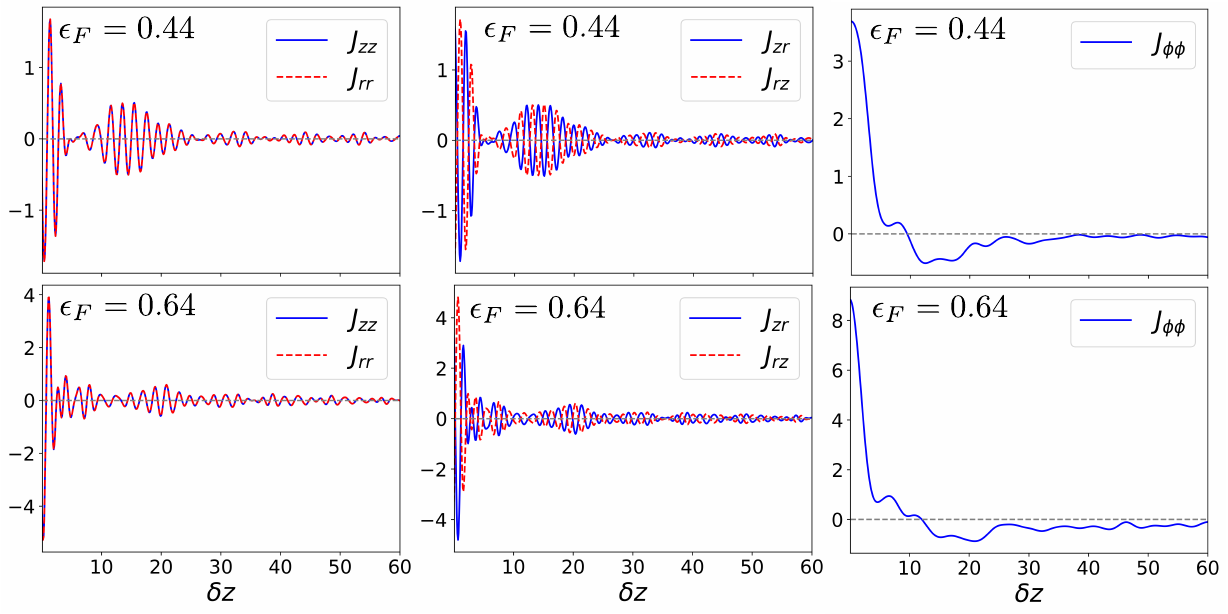}
	\caption{We plot all the distinct effective spin-spin correlation between impurity spins mediated by conduction electrons on Weyl-semimetal nano-wire as a function of $\delta z$ for fixed $\delta \phi$ in cylindrical coordinate ($\delta z=z_{1}-z_{2}, \delta \phi=\phi_{1}-\phi_{2}$). The symmetries found from the numerical simulation are as follows: $J_{zz}=J_{rr}$, $J_{rz}=-J_{zr}$, $J_{r \phi}=-J_{\phi r}$, $J_{z \phi}=J_{\phi z}$. The magnitude of the correlations $J_{r \phi}, J_{\phi r}, J_{z \phi}, J_{\phi z}$ are very small compared to the other correlations that are shown here. Here, the RKKY couplings depend on three frequencies which are determined primarily by two scales, namely the Fermi energy and the width of each Fermi arc. Also, in the parameter range used here, the coupling $J_{\phi \phi}$ is predominately antiferromagnetic, ferromagnetic for small and large $\delta z$ respectively. On the top panel the Fermi energy is set at $\epsilon_{F}=0.44$, and on the bottom panel $\epsilon_{F}=0.64$. The Fermi energy is set to values that include contributions from both Fermi arcs and low-energy bulk states. Other parameters are as follows : $R=7$, $\delta \phi=0.1$, $k_{0}=0.05 \pi$, $|m_{\text{max}}|=9$. All the correlations are measured in the unit of $10^{-3} J^{2}$. Numerical convergence are upto order $10^{-7} J^{2}$.} 
	\label{wsmcorr1}
\end{figure*}

In Fig~(\ref{fig:wsmnewdispersion})(a), we show the energy dispersion as a function of $k_{z}$ for the Weyl-semimetal nanowire. The gap in the low-energy flat bands is inversely proportional to the radius of the cylinder. The flatness of Fermi arc surface strongly depends on the boundary condition at the surface of the cylinder~\cite{PhysRevB.104.155425} (here we choose infinite mass boundary condition).  In Fig~(\ref{fig:wsmnewdispersion})(b) the spin densities of the Fermi-arc surface states are plotted with respect to the radial direction ($r$) which shows that there is a spin momentum locking along $\hat{\phi}$ direction. 

\section{RKKY interaction}\label{rkkyanalytic}

We place two impurity spins on the surface of a WSM nanowire (NW) described by Eq.(\ref{eq:Hamiltonian}). Ruderman-Kittel-Kasuya-Yosida (RKKY) interaction accounts for the effective coupling between these two spins. The coupling between magnetic impurities and conduction electron is governed by $s-d$ interaction~\cite{RevModPhys.78.809},

\begin{equation}
H_{s-d}=J\sum_{i=1,2} {\bf S}_{i} \cdot {\bf \sigma} \delta(\bf{r}-\bf{R}_{i}).
\end{equation}
$J$ stands for the strength of $s-d$  interaction and $\bf{R}_{i}$ being the positions of the magnetic impurity. RKKY interaction between two spin at position $(R,\phi,z)$ and $(R,\phi',z')$ due to indirect exchange interaction is given by~\cite{chang2015rkky},
\begin{align}\label{RKKY}
\nonumber
  &H_{RKKY} \\  &=  -\frac{J^{2}}{\pi} \Im{ \int_{-\infty}^{\epsilon_{F}} d\omega  \rm{Tr}[{\bf S}_{1}({\bf r}).\sigma_{\bf r} \textit{G} ({\bf r },{\bf r}',\omega+i \eta){\bf S}_{2}({\bf r}').\sigma_{\bf r'}}],\\
  \nonumber
&G({\bf r }',{\bf r},\omega+i \eta)]\\
&
\equiv\sum_{l,k=x,y,z} J_{lk} S_{1l}S_{2k}.  
\end{align}

Here, $G({\bf r },{\bf r}',\omega+i \eta)$ is the real-space Green’s function for the unperturbed electron system, $\epsilon_{F}$ is the Fermi energy. $J_{lk}$ is the electron spin-spin correlation matrix. The last line of the equation is the effective spin Hamiltonian between two different impurity spins where the interactions are mediated by the Weyl electrons.

\subsection{Low energy effective spin model for WSM}

All the details of a specific electronic system in which the impurity spin is being embedded are entered into the calculations via Green's function $G({\bf r },{\bf r}')$. We consider the effective low-energy model for the WSM nanowires, considering only the Fermi-arc surface states and a few bulk states. In general, Green's function in its spectral decomposition form can be written as~\cite{coleman2017introduction},

\begin{equation}\label{spectraldecom}
\begin{aligned}
G({\bf r },{\bf r}',\omega+i \eta)&=\dfrac{1}{V} \sum_{m,k,\xi}\frac{\psi_{m,k,\xi}({\bf r})\psi_{m,k,\xi}^{\dagger}({\bf  r}')}{\omega-\epsilon_{m,k,\xi}+i\eta},
\end{aligned}
\end{equation}
with $\epsilon_{m,k,\xi}$ and $\psi_{m,k,\xi}$ are the dispersion and wavefunction of the $m^{th}$ band of chirality $\xi$ at momentum $k=k_{z}$ respectively.
From Eq.~\eqref{wavefunctionwsm} and Eq.~(\ref{transdental2}) we can write the solution of the wavefunction as (with appropriate renormalization constant)
\begin{equation}\label{eigenfunc}
\psi_{m,k,\xi}({\bf r})=e^{i m \phi} e^{i k z}\begin{pmatrix}
u_{m,k,\xi} \\v_{m,k,\xi}e^{i\phi}
\end{pmatrix}.
\end{equation}
By substitution of Eq.~(\ref{eigenfunc}) in the Eq.~(\ref{spectraldecom}) we get the form of the Green's function,

\begin{equation}\label{wsmGfunction}
\begin{aligned}
G({\bf r},{\bf  r}',\omega+i \eta)&=\int_{\bf k} \sum_{m,\xi} \frac{e^{i(m\delta \phi+k\delta z)}}{\omega-\epsilon_{m,\xi,k}+i\eta}
\\&~~~~~
\begin{pmatrix}
|u_{m,k,\xi}|^{2}&  u_{m,k,\xi} v_{m,k,\xi}^{*} e^{-i \phi'} \\ v_{m,k,\xi} u_{m,k,\xi}^{*} e^{i \phi} & |v_{m,k,\xi}|^{2}e^{i\delta \phi},
\end{pmatrix},
\end{aligned}
\end{equation}
where we denote $\delta \phi= (\phi-\phi')$ and $\delta z= (z- z')$, $\phi/\phi'$ and $z,z'$ is the azimuthal and longitudinal coordinates of the first and second impurity spins respectively. Four Weyl nodes are located at $k_{z}=\pm k_{0} \pm \pi/2$ . 

In Appendix~\ref{app}, we provide the details of the RKKY correlations for the low energy states in cylindrical coordinate. The corresponding expressions for the rectangular coordinate can be obtained just by usual coordinate transformation. The analytical expressions used to calculate different RKKY couplings numerically can be found in the in Appendix~\ref{app}(Eq.~\eqref{EqApp1}-\eqref{EqApp9}).

\begin{figure*}
    \centering\includegraphics[width=1.0\linewidth]{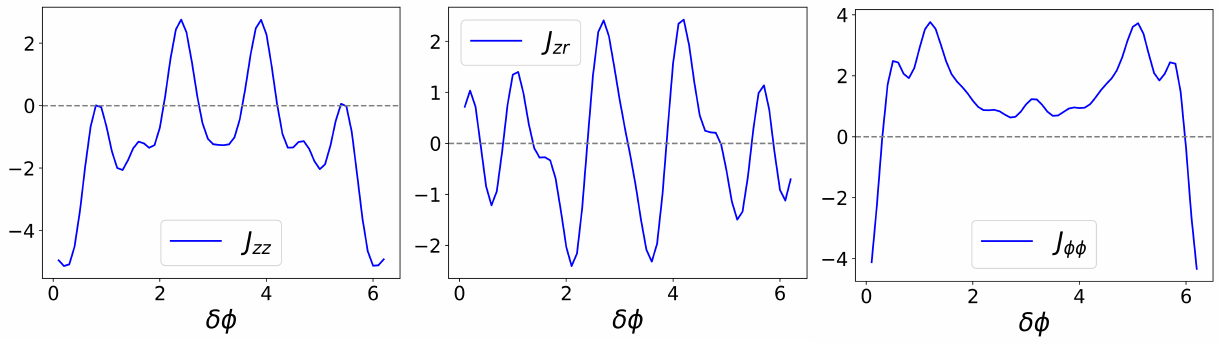}
    \caption{In this figure we have compared the distinct RKKY correlations as a function of $\delta \phi$ for fixed $\delta z=2.0$, $\epsilon_{F}=0.44$. Other parameters are the same as Fig.~(\ref{wsmcorr1}). RKKY correlation $J_{zz}$ and $J_{\phi \phi}$ symmetric around $\delta \phi =\pi$ but, $J_{zr}$ shows antisymmetric dependence. All the correlations are measured in the unit of $10^{-3} J^{2}$. Numerical convergence are atleast $10^{-7} J^{2}$.}
    \label{fig:my_label_small2}
\end{figure*}

\subsection{Contribution explicitly from Fermi arc surface states}
If we only take into account  the RKKY interactions which are mediated by the Fermi-arc surface states, then
the wavefunction in Eq.~\eqref{eigenfunc} can be approximated by~\cite{sonurkkyqsh}, 

\begin{equation}\label{Eq:FAwave}
\psi_{m,k,\xi}({\bf r}) \propto e^{i m \phi} e^{i k z}\begin{pmatrix}1
  \\-i\xi e^{i\phi}
\end{pmatrix}, 
\end{equation}
with dispersion,
\begin{equation}\label{Eq:FAdis}
   \epsilon_{m,k,\xi}=(m+ \dfrac{1}{2})\ \xi \dfrac{v_{F}}{R}.  
\end{equation}
 For the Fermi-arc surface states the energy  dispersion is independent of $k_z$. The corresponding Green's function is given by, (upto a normalization constant)
\begin{equation}\label{green-fermi}
\begin{aligned}
G({\bf r},{\bf  r}') & \propto \dfrac{1}{V} \sum_{m,k,\xi} \frac{e^{i(m\delta \phi+k\delta z)}}{\omega-(m+ \dfrac{1}{2})\ \xi \dfrac{v_{F}}{R}+i\eta}
\\&~~~~~
\begin{pmatrix}
1 & -i\xi e^{i \phi} \\ i\xi e^{-i \phi'} & e^{i \delta \phi}
\end{pmatrix}
\theta(k-k_{\xi,1}) \theta(k_{\xi,2}-k).
\end{aligned}
\end{equation}

Note that, $\eta$ is a small parameter that introduces a cutoff in the integral with, $\eta=1/ \tau_{F}$, where $\tau_{F}$ is the lifetime of FA surface states. $v_{F} \eta=\lambda_{mf}$, with $\lambda_{mf}$, being the mean-free path of the FA surface states, in order for the impurity spins to interact via RKKY coupling, their separation should be small compared than $\lambda_{mf}$~\cite{sonurkkyqsh}. The theta functions introduced in Eq.~(\ref{green-fermi}) serve as a cutoff for the momentum integral. $k_{\xi,1/2}$ represents the location of four Weyl nodes with $\xi=\pm 1$, also we have replaced $k_z=k$. We are now in a position to discuss the detailed symmetries of the effective RKKY spin Hamiltonian and our numerical findings in the following section.\\

\section{Results}\label{Sec:numericalresults}
\subsection{Results for the low energy Bulk states}

When the Fermi energy is set within the low energy bulk bands, the dominant contribution for the RKKY correlations comes from the WSM bulk states. In Fig.~(\ref{wsmcorr1}) we plot different RKKY correlation matrix elements as a function of $\delta z= z_{1}-z_{2}$ ($z_{1},z_{2}$ being the vertical position of the first and second impurity spins respectively) for fixed $\delta \phi$ ($\phi$ is the azimuthal position). Out of nine correlations only five are of similar magnitude and the rest are very small compared to them (see  Appendix~\ref{app} for more detail). The effective low energy spin Hamiltonian between the impurities can be written in the following form:

\begin{widetext}
    \begin{equation}\label{wsmeffspinH}
H_{EF}=\sum_{\alpha=r,\phi,z} J_{\alpha\alpha}S^{1}_{\alpha}S^{2}_{\alpha}+D_{\phi}\hat{\phi} \cdot (\vec{S^{1}} \times \vec{S^{2}})+D_{z}\hat{z} \cdot (\vec{S^{1}} \times \vec{S^{2}})+\Gamma_{r} (S^{1}_{z}S^{2}_{\phi}+S^{1}_{\phi}S^{2}_{z}),
    \end{equation} 
\end{widetext}
with, $J_{rr}=J_{zz}$, $D_{\phi}=J_{rz}=-J_{zr}$, $D_{z}=J_{r\phi}=-J_{\phi r}$, $\Gamma_{r}=J_{z\phi}=J_{\phi z}$. The first term is anisotropic exchange interaction, the second and third terms are DMI-type coupling and the last term is symmetric off-diagonal exchange interaction. Although, from numerical simulations the magnitude of the last two terms is very small compared to the first two terms. Predominately the RKKY interaction for the low energy bulk state scale as $\delta z^{-\eta}$, where $1<\eta<2$. From Fig.~(\ref{wsmcorr1}) it is clear that the correlation $J_{zz} (=J_{rr})$ and $J_{zr}(=-J_{rz})$ has two oscillation frequency whereas $J_{\phi\phi}$ has only a single frequency.
Also, among all the correlations, the coupling $J_{\phi \phi}$ is largest in magnitude at small distance. In general when the Fermi energy is set at non-zero value we expect the usual $2k_{F}$ oscillation in the RKKY coupling, where $k_{F}$ is the Fermi wave vector. When the Fermi energy is increased from $\epsilon_{F}=0.44$ to $\epsilon_{F}=0.64$ the oscillation frequency increases for $J_{zz}$ and $J_{zr}$. The magnitude of all the correlations increases with increasing Fermi energy. The envelope within which these oscillations occur should behave according the case when we set Fermi energy at zero which we discuss in the next section. We also plot the RKKY correlations for fixed $\delta z$ as a function of $\delta \phi$ in Fig~(\ref{fig:my_label_small2}).

\subsection{Results for the Surface states}
Now we focus explicitly the contributions coming from the Fermi arc surface states by calculating the RKKY correlations using Eq~(\ref{Eq:FAwave})-(\ref{Eq:FAdis}).  For the Fermi arc surface states the correlation $J_{\phi z}=J_{z\phi}=J_{r\phi}=J_{\phi r}=0$ becomes exactly zero. Symmetries for the rest of the correlations remain the same as the bulk states. In the following, we present the exact analytical expression for the non-zero distinct RKKY couplings when the Fermi energy is set at $\epsilon_{F}=0$.
\begin{widetext}
	
%\begin{tcolorbox}[colback=white!5!white,colframe=black!90!black]	
\begin{equation}
\begin{aligned}
	J_{zz}=&\frac{-J^{2}}{\pi} \Im \int_{\omega=-\infty}^{\omega=\epsilon_{F}} [G^{++}({\bf r },{\bf r}')G^{++}({\bf r }',{\bf r})-G^{+-}({\bf r },{\bf r}')G^{-+}({\bf r }',{\bf r})
	-G^{-+}({\bf r },{\bf r}')G^{+-}({\bf r }',{\bf r})+G^{--}({\bf r },{\bf r}')G^{--}({\bf r }',{\bf r})]\\
	& \propto J^{2}(\Delta k)^2\sum_{m,k_{\xi},\xi,n,k'_{\xi'},\xi'}\cos \left[(m-n)\delta \phi+(k_{\xi}-k'_{\xi'})\delta z\right]\dfrac{\Big[- \sgn[\epsilon_{m,\xi}]+\sgn[\epsilon_{n,\xi'}]] \Big]}{\epsilon_{m,\xi}-\epsilon_{n,\xi'}}(1-\xi \xi')\\
	&=-16 J^{2}\sum_{m,n}\Big(\dfrac{\sin[k_{0}\delta z]}{\delta z}\Big)^{2}\cos[\pi \delta z+(m-n)\delta \phi]\dfrac{\Big[ \sgn[\epsilon_{m,1}]+\sgn[\epsilon_{n,1}]] \Big]}{\dfrac{v_{F}}{R}(m+n+1)},\\ 
\end{aligned}
\end{equation}
%\end{tcolorbox}

\begin{equation}
\begin{aligned}
J_{\phi\phi}&=\frac{-J^{2}}{\pi} \Im \int_{-\infty}^{\epsilon_{F}} d\omega [-G^{-+}({\bf r },{\bf r}')G^{-+}({\bf r }',{\bf r})e^{-i(\phi+\phi')}+G^{--}({\bf r },{\bf r}')G^{++}({\bf r }',{\bf r})e^{-i\delta \phi}
+e^{-i\delta \phi}G^{++}({\bf r },{\bf r}')G^{--}({\bf r }',{\bf r})\\
& \propto J^{2}(\Delta k)^2\sum_{m,k_{\xi},\xi,n,k'_{\xi'},\xi'}\cos \left[(m-n)\delta \phi+(k_{\xi}-k'_{\xi'})\delta z\right]\dfrac{\Big[-\sgn[\epsilon_{m,\xi}]+\sgn[\epsilon_{n,\xi'}]] \Big]}{\epsilon_{m,\xi}-\epsilon_{n,\xi'}}(1+\xi \xi')\\
&=-16 J^{2}\sum_{m,n}\Big(\dfrac{\sin[k_{0}\delta z]}{\delta z}\Big)^{2}\cos[(m-n)\delta \phi]\dfrac{\Big[-\sgn[\epsilon_{m,1}]+\sgn[\epsilon_{n,1}]] \Big]}{\dfrac{v_{F}}{R}(m-n)}, 
\end{aligned}
\end{equation}
\\
\begin{equation}
\begin{aligned}
J_{zr}&=\frac{-J^{2}}{\pi} \Im \int_{-\infty}^{\epsilon_{F}} d\omega [e^{-i\phi'}G^{++}({\bf r },{\bf r}')G^{-+}({\bf r }',{\bf r})+e^{i\phi'}G^{+-}({\bf r },{\bf r}')G^{++}({\bf r }',{\bf r})-e^{i\phi'}G^{--}({\bf r },{\bf r}')G^{+-}({\bf r }',{\bf r})-\\
&e^{-i\phi'}G^{-+}({\bf r },{\bf r}')G^{--}({\bf r }',{\bf r})]\\
& \propto J^{2}(\Delta k)^2\sum_{m,k_{\xi},\xi,n,k'_{\xi'},\xi'}\sin \left[(m-n)\delta \phi+(k_{\xi}-k'_{\xi'})\delta z\right]
\dfrac{-\sgn\Big[\epsilon_{m,\xi}\Big]+\sgn\Big[\epsilon_{n,\xi}\Big]}{\epsilon_{m,\xi}-\epsilon_{n,\xi'}}
\Big[-\xi +\xi'\Big]\\
&=-16 J^{2}\sum_{m,n}\Big(\dfrac{\sin[k_{0}\delta z]}{\delta z}\Big)^{2}\sin[\pi \delta z+(m-n)\delta \phi]\dfrac{\Big[- \sgn[\epsilon_{m,1}]+\sgn[\epsilon_{n,-1}]] \Big]}{\dfrac{v_{F}}{R}(m+n+1)},
\end{aligned}
\end{equation}
\\
\end{widetext}

Where $(\Delta k)^2=\dfrac{4\pi^2}{N_{k}^2}$ ($\int_{\-\pi}^{\pi}  dk_{z}=\sum_{i=1}^{N_{k}}  \Delta k$). We want to emphasize several important points here. Firstly, all the above correlations have $(\sin{(k_{0}\delta z)}/\delta z)^2$ dependence. This indicates the \textit{two-dimensional} signatures of the Fermi-arc states.  Secondly, The frequency of RKKY oscillations depends on the width of each Fermi-arc ($2k_{0}$). Thirdly, the correlation $J_{zz}$ and $J_{zr}$ are only nonzero when $\xi'=\bar{\xi}$ whereas $J_{\phi \phi}$ is only nonzero when $\xi'=\xi$.
The symmetries between the elements of the correlation matrix described in the preceding paragraph remain intact as in Fig.~(\ref{wsmcorr1}). 

\subsection{Discussion on the symmetries and magnitudes of the effective coupling:}
Now we discuss the symmetry structure of various spin spin correlations coming from Fermi-arc surface states as well as the low energy bulk states. 
The symmetry structure, when only the contributions from the Fermi arcs are considered, is largely dictated by the cylindrical geometry and the spin-momentum locking of the Fermi-arc surface states. The interactions $J_{rr}=J_{zz}$ are equal, which is due to the polarization of the Fermi-arc surface states along the $\hat{\phi}$ (azimuthal) direction. As a result, the couplings in the directions \textit{orthogonal} to $\hat{\phi}$, i.e., along $\hat{r}$ and $\hat{z}$, are equivalent. The anti-symmetric term $J_{rz}=-J_{zr}$ primarily arises from the chiral nature of the Fermi-arc states, as the spin orientations are locked with the motion of the electrons. When these conduction electrons mediate the RKKY interactions between impurity spins, the interaction depends on the relative positions of the impurities and the surface electron spins. Consequently, this leads to the Dzyaloshinskii-Moriya interaction (DMI). Also, since the Fermi arc surface states are polarized along the $\hat{\phi}$ direction, the RKKY interaction is expected to be strongest when both impurity spins are also aligned along $\hat{\phi}$, which is evident in Fig~\ref{wsmcorr1}. Also, the correlation \(J_{\phi\phi}\) is distinct from \(J_{zz}\) and \(J_{rz}\). 
In particular, we do not observe the rapid \(2k_F\)-type oscillations that arise 
from scattering processes connecting states at \(+k_F\) and \(-k_F\). 
Due to the spin--momentum locking of the Fermi-arc surface states, 
\(J_{\phi\phi}\) is nonzero only when \(\xi = \xi'\). 
As a result, it cannot mediate scattering between states at \(+k_z\) and \(-k_z\).

Moreover, the interactions $J_{z\phi}=J_{\phi z}=J_{r\phi}=J_{\phi r}=0$. In this case, no RKKY interactions couple $\hat{\phi}$ with any other directions, which indicates, there is no transverse spin exchange between the $\phi$ component and the $r$ or $z$ components. An alternative way to understand this is that the Fermi arc surface states are eigenstates of $\sigma_{\phi}$. In the RKKY correlation, the matrix elements for $\sigma_{\phi}$ are nonzero only when $\xi = \xi'$, whereas for $\sigma_{z}$ and $\sigma_{r}$ the nonzero matrix elements occur when $\xi = -\xi'$. As a result, the correlations $J_{x\phi}$ (with $x = r, z$) vanish. The same selection-rule argument can be extended to explain why the other components are nonzero.
\\

However, when the Fermi energy is set to a value such that contributions from the low energy bulk states are also included, the electron polarizations are no longer purely along the $\phi$ direction. While the overall symmetry remains intact, the other off-diagonal components acquire small but nonzero values, with the following symmetry: $J_{z\phi}=J_{\phi z}$ and $J_{r\phi}=-J_{\phi r}$. The symmetry $J_{z\phi}=J_{\phi z}$ arises due to the cylindrical geometry, which enforces a cylindrical symmetry along the $z$ direction, leading to symmetric interactions. In contrast, $J_{r\phi}=-J_{\phi r}$ is antisymmetric because the radial direction breaks the inversion symmetry. So our numerical, as well as our numerical results suggest that the symmetry of the RKKY correlations for the surface as well as the low energy bulk states are strongly dictated by the nature of the Fermi arcs.
\\

Finally, we examine the scaling of the RKKY coupling with $\delta z$ for a fixed $\delta \phi$. For the case of Fermi-arc surface states, the coupling follows a $1/(\delta z)^2$ dependence. In contrast, when low-energy bulk contributions are included, the decay becomes slower, following $1/(\delta z)^{\eta}$ with $\eta<2$, particularly for small cylinder radii. The faster decay observed for surface-state-mediated couplings arises from the non-dispersive nature and finite extent of the Fermi arcs. However, when both low-energy bulk states and surface states are present, the decay slows down, as confirmed by our numerical results. For alternative mass boundary conditions, where dispersive Fermi-arc states emerge, increasing the sample radius can lead to a competition between Fermi-arc surface states and bulk states. In such situations, the finite lifetime of the Fermi-arc states may also become an important factor.\\

\section{SUMMARY and OUTLOOK}\label{discussion}
In this work, we studied the RKKY interaction between impurity spins located on the surface of a Weyl semimetal (WSM) nanowire. Our analysis reveals that in such a system, the effective RKKY interaction comprises anisotropic Heisenberg-type terms, the Dzyaloshinskii–Moriya (DM) interaction, and symmetric off-diagonal spin–spin components. We find that the Fermi-arc surface states play a crucial role in mediating the RKKY coupling, beyond the bulk contribution. When only the Fermi-arc surface states are taken into account, the RKKY interaction decays as $1/(\delta z)^2$.  In contrast, when both the bulk states and the Fermi-arc surface states are included, the decay is slower, with $1/(\delta z)^{\eta}$ where $\eta<2$. This indicates even longer-range correlations. Nevertheless, the symmetry structure of the RKKY correlation matrix remains the same for both the surface-state and bulk states, a property determined primarily by the cylindrical geometry and the mass boundary condition.\\
 
 The sign and magnitude of these interactions can be tuned by altering the relative positions of the impurity spins. One can think of experimentally feasible studies based on our work. The recent advances in engineering techniques offer a wide range of possibilities, from quantum transport to magnetic or spectroscopic measurements. For example, the anisotropic Heisenberg exchange term can be probed via SQUID magnetometry by measuring the magnetic susceptibility~\cite{mugiraneza2022tutorial}, while the DM interaction may be detected using magneto-optical Kerr microscopy~\cite{PhysRevLett.119.077205}.
In particular, the interplay between the DM and Heisenberg interactions can lead to the emergence of skyrmions and other non-trivial spin textures~\cite{PhysRevB.109.L201108}, which may have promising applications in quantum information storage. Our findings present an alternative pathway for engineering topological semimetals, with potential applications in spintronics and optics. Furthermore, tuning the symmetric off-diagonal interaction terms can induce anisotropic permittivity and distinctive optical responses, making the Weyl semimetal a suitable candidate for optical isolator applications~\cite{wu2023tunable}.
The recent successful synthesis of Weyl semimetal NbAs nanowires~\cite{cheon2025surface} through thermomechanical nanomodeling offers an exciting avenue for engineering novel spin Hamiltonians mediated by RKKY interactions in WSMs in the future.

\section{Acknowledgment}
RM and AD acknowledge R.Kundu and S.Verma for their useful comments and suggestions. R.M. acknowledges the CSIR (Govt. of India) for financial support.

\renewcommand{\thefigure}{A\arabic{figure}}
\setcounter{figure}{0}
\renewcommand{\theequation}{A\arabic{equation}}
\setcounter{equation}{0}
\bibliography{main.bib}

\begin{thebibliography}{57}%
\makeatletter
\providecommand \@ifxundefined [1]{%
 \@ifx{#1\undefined}
}%
\providecommand \@ifnum [1]{%
 \ifnum #1\expandafter \@firstoftwo
 \else \expandafter \@secondoftwo
 \fi
}%
\providecommand \@ifx [1]{%
 \ifx #1\expandafter \@firstoftwo
 \else \expandafter \@secondoftwo
 \fi
}%
\providecommand \natexlab [1]{#1}%
\providecommand \enquote  [1]{``#1''}%
\providecommand \bibnamefont  [1]{#1}%
\providecommand \bibfnamefont [1]{#1}%
\providecommand \citenamefont [1]{#1}%
\providecommand \href@noop [0]{\@secondoftwo}%
\providecommand \href [0]{\begingroup \@sanitize@url \@href}%
\providecommand \@href[1]{\@@startlink{#1}\@@href}%
\providecommand \@@href[1]{\endgroup#1\@@endlink}%
\providecommand \@sanitize@url [0]{\catcode `\\12\catcode `\$12\catcode `\&12\catcode `\#12\catcode `\^12\catcode `\_12\catcode `\%12\relax}%
\providecommand \@@startlink[1]{}%
\providecommand \@@endlink[0]{}%
\providecommand \url  [0]{\begingroup\@sanitize@url \@url }%
\providecommand \@url [1]{\endgroup\@href {#1}{\urlprefix }}%
\providecommand \urlprefix  [0]{URL }%
\providecommand \Eprint [0]{\href }%
\providecommand \doibase [0]{https://doi.org/}%
\providecommand \selectlanguage [0]{\@gobble}%
\providecommand \bibinfo  [0]{\@secondoftwo}%
\providecommand \bibfield  [0]{\@secondoftwo}%
\providecommand \translation [1]{[#1]}%
\providecommand \BibitemOpen [0]{}%
\providecommand \bibitemStop [0]{}%
\providecommand \bibitemNoStop [0]{.\EOS\space}%
\providecommand \EOS [0]{\spacefactor3000\relax}%
\providecommand \BibitemShut  [1]{\csname bibitem#1\endcsname}%
\let\auto@bib@innerbib\@empty
%</preamble>
\bibitem [{\citenamefont {Armitage}\ \emph {et~al.}(2018)\citenamefont {Armitage}, \citenamefont {Mele},\ and\ \citenamefont {Vishwanath}}]{armitage2018weyl}%
  \BibitemOpen
  \bibfield  {author} {\bibinfo {author} {\bibfnamefont {N.}~\bibnamefont {Armitage}}, \bibinfo {author} {\bibfnamefont {E.}~\bibnamefont {Mele}},\ and\ \bibinfo {author} {\bibfnamefont {A.}~\bibnamefont {Vishwanath}},\ }\bibfield  {title} {\bibinfo {title} {Weyl and dirac semimetals in three-dimensional solids},\ }\href {https://journals.aps.org/rmp/abstract/10.1103/RevModPhys.90.015001} {\bibfield  {journal} {\bibinfo  {journal} {Reviews of Modern Physics}\ }\textbf {\bibinfo {volume} {90}},\ \bibinfo {pages} {015001} (\bibinfo {year} {2018})}\BibitemShut {NoStop}%
\bibitem [{\citenamefont {Jia}\ \emph {et~al.}(2016)\citenamefont {Jia}, \citenamefont {Xu},\ and\ \citenamefont {Hasan}}]{jia2016weyl}%
  \BibitemOpen
  \bibfield  {author} {\bibinfo {author} {\bibfnamefont {S.}~\bibnamefont {Jia}}, \bibinfo {author} {\bibfnamefont {S.-Y.}\ \bibnamefont {Xu}},\ and\ \bibinfo {author} {\bibfnamefont {M.~Z.}\ \bibnamefont {Hasan}},\ }\bibfield  {title} {\bibinfo {title} {Weyl semimetals, fermi arcs and chiral anomalies},\ }\href {https://www.nature.com/articles/nmat4787} {\bibfield  {journal} {\bibinfo  {journal} {Nature materials}\ }\textbf {\bibinfo {volume} {15}},\ \bibinfo {pages} {1140} (\bibinfo {year} {2016})}\BibitemShut {NoStop}%
\bibitem [{\citenamefont {Rao}(2016)}]{rao2016weyl}%
  \BibitemOpen
  \bibfield  {author} {\bibinfo {author} {\bibfnamefont {S.}~\bibnamefont {Rao}},\ }\bibfield  {title} {\bibinfo {title} {Weyl semi-metals: a short review},\ }\href {https://arxiv.org/abs/1603.02821} {\bibfield  {journal} {\bibinfo  {journal} {arXiv preprint arXiv:1603.02821}\ } (\bibinfo {year} {2016})}\BibitemShut {NoStop}%
\bibitem [{\citenamefont {Burkov}\ and\ \citenamefont {Balents}(2011)}]{PhysRevLett.107.127205}%
  \BibitemOpen
  \bibfield  {author} {\bibinfo {author} {\bibfnamefont {A.~A.}\ \bibnamefont {Burkov}}\ and\ \bibinfo {author} {\bibfnamefont {L.}~\bibnamefont {Balents}},\ }\bibfield  {title} {\bibinfo {title} {Weyl semimetal in a topological insulator multilayer},\ }\href {https://doi.org/10.1103/PhysRevLett.107.127205} {\bibfield  {journal} {\bibinfo  {journal} {Phys. Rev. Lett.}\ }\textbf {\bibinfo {volume} {107}},\ \bibinfo {pages} {127205} (\bibinfo {year} {2011})}\BibitemShut {NoStop}%
\bibitem [{\citenamefont {Chang}\ \emph {et~al.}(2016)\citenamefont {Chang}, \citenamefont {Xu}, \citenamefont {Zheng}, \citenamefont {Lee}, \citenamefont {Huang}, \citenamefont {Belopolski}, \citenamefont {Sanchez}, \citenamefont {Bian}, \citenamefont {Alidoust}, \citenamefont {Chang}, \citenamefont {Hsu}, \citenamefont {Jeng}, \citenamefont {Bansil}, \citenamefont {Lin},\ and\ \citenamefont {Hasan}}]{arca}%
  \BibitemOpen
  \bibfield  {author} {\bibinfo {author} {\bibfnamefont {G.}~\bibnamefont {Chang}}, \bibinfo {author} {\bibfnamefont {S.-Y.}\ \bibnamefont {Xu}}, \bibinfo {author} {\bibfnamefont {H.}~\bibnamefont {Zheng}}, \bibinfo {author} {\bibfnamefont {C.-C.}\ \bibnamefont {Lee}}, \bibinfo {author} {\bibfnamefont {S.-M.}\ \bibnamefont {Huang}}, \bibinfo {author} {\bibfnamefont {I.}~\bibnamefont {Belopolski}}, \bibinfo {author} {\bibfnamefont {D.~S.}\ \bibnamefont {Sanchez}}, \bibinfo {author} {\bibfnamefont {G.}~\bibnamefont {Bian}}, \bibinfo {author} {\bibfnamefont {N.}~\bibnamefont {Alidoust}}, \bibinfo {author} {\bibfnamefont {T.-R.}\ \bibnamefont {Chang}}, \bibinfo {author} {\bibfnamefont {C.-H.}\ \bibnamefont {Hsu}}, \bibinfo {author} {\bibfnamefont {H.-T.}\ \bibnamefont {Jeng}}, \bibinfo {author} {\bibfnamefont {A.}~\bibnamefont {Bansil}}, \bibinfo {author} {\bibfnamefont {H.}~\bibnamefont {Lin}},\ and\ \bibinfo {author} {\bibfnamefont {M.~Z.}\ \bibnamefont {Hasan}},\ }\bibfield  {title} {\bibinfo {title}
  {Signatures of fermi arcs in the quasiparticle interferences of the weyl semimetals taas and nbp},\ }\href {https://doi.org/10.1103/PhysRevLett.116.066601} {\bibfield  {journal} {\bibinfo  {journal} {Phys. Rev. Lett.}\ }\textbf {\bibinfo {volume} {116}},\ \bibinfo {pages} {066601} (\bibinfo {year} {2016})}\BibitemShut {NoStop}%
\bibitem [{\citenamefont {Gyenis}\ \emph {et~al.}(2016)\citenamefont {Gyenis}, \citenamefont {Inoue}, \citenamefont {Jeon}, \citenamefont {Zhou}, \citenamefont {Feldman}, \citenamefont {Wang}, \citenamefont {Li}, \citenamefont {Jiang}, \citenamefont {Gibson}, \citenamefont {Kushwaha}, \citenamefont {Krizan}, \citenamefont {Ni}, \citenamefont {Cava}, \citenamefont {Bernevig},\ and\ \citenamefont {Yazdani}}]{arcb}%
  \BibitemOpen
  \bibfield  {author} {\bibinfo {author} {\bibfnamefont {A.}~\bibnamefont {Gyenis}}, \bibinfo {author} {\bibfnamefont {H.}~\bibnamefont {Inoue}}, \bibinfo {author} {\bibfnamefont {S.}~\bibnamefont {Jeon}}, \bibinfo {author} {\bibfnamefont {B.~B.}\ \bibnamefont {Zhou}}, \bibinfo {author} {\bibfnamefont {B.~E.}\ \bibnamefont {Feldman}}, \bibinfo {author} {\bibfnamefont {Z.}~\bibnamefont {Wang}}, \bibinfo {author} {\bibfnamefont {J.}~\bibnamefont {Li}}, \bibinfo {author} {\bibfnamefont {S.}~\bibnamefont {Jiang}}, \bibinfo {author} {\bibfnamefont {Q.~D.}\ \bibnamefont {Gibson}}, \bibinfo {author} {\bibfnamefont {S.~K.}\ \bibnamefont {Kushwaha}}, \bibinfo {author} {\bibfnamefont {J.~W.}\ \bibnamefont {Krizan}}, \bibinfo {author} {\bibfnamefont {N.}~\bibnamefont {Ni}}, \bibinfo {author} {\bibfnamefont {R.~J.}\ \bibnamefont {Cava}}, \bibinfo {author} {\bibfnamefont {B.~A.}\ \bibnamefont {Bernevig}},\ and\ \bibinfo {author} {\bibfnamefont {A.}~\bibnamefont {Yazdani}},\ }\bibfield  {title} {\bibinfo {title} {Imaging
  electronic states on topological semimetals using scanning tunneling microscopy},\ }\href {https://doi.org/10.1088/1367-2630/18/10/105003} {\bibfield  {journal} {\bibinfo  {journal} {New Journal of Physics}\ }\textbf {\bibinfo {volume} {18}},\ \bibinfo {pages} {105003} (\bibinfo {year} {2016})}\BibitemShut {NoStop}%
\bibitem [{\citenamefont {Huang}\ \emph {et~al.}(2015{\natexlab{a}})\citenamefont {Huang}, \citenamefont {Xu}, \citenamefont {Belopolski}, \citenamefont {Lee}, \citenamefont {Chang}, \citenamefont {Wang}, \citenamefont {Alidoust}, \citenamefont {Bian}, \citenamefont {Neupane}, \citenamefont {Zhang} \emph {et~al.}}]{arcc}%
  \BibitemOpen
  \bibfield  {author} {\bibinfo {author} {\bibfnamefont {S.-M.}\ \bibnamefont {Huang}}, \bibinfo {author} {\bibfnamefont {S.-Y.}\ \bibnamefont {Xu}}, \bibinfo {author} {\bibfnamefont {I.}~\bibnamefont {Belopolski}}, \bibinfo {author} {\bibfnamefont {C.-C.}\ \bibnamefont {Lee}}, \bibinfo {author} {\bibfnamefont {G.}~\bibnamefont {Chang}}, \bibinfo {author} {\bibfnamefont {B.}~\bibnamefont {Wang}}, \bibinfo {author} {\bibfnamefont {N.}~\bibnamefont {Alidoust}}, \bibinfo {author} {\bibfnamefont {G.}~\bibnamefont {Bian}}, \bibinfo {author} {\bibfnamefont {M.}~\bibnamefont {Neupane}}, \bibinfo {author} {\bibfnamefont {C.}~\bibnamefont {Zhang}}, \emph {et~al.},\ }\bibfield  {title} {\bibinfo {title} {A weyl fermion semimetal with surface fermi arcs in the transition metal monopnictide taas class},\ }\href {https://www.nature.com/articles/ncomms8373} {\bibfield  {journal} {\bibinfo  {journal} {Nature communications}\ }\textbf {\bibinfo {volume} {6}},\ \bibinfo {pages} {7373} (\bibinfo {year}
  {2015}{\natexlab{a}})}\BibitemShut {NoStop}%
\bibitem [{\citenamefont {Xu}\ \emph {et~al.}(2016)\citenamefont {Xu}, \citenamefont {Belopolski}, \citenamefont {Sanchez}, \citenamefont {Neupane}, \citenamefont {Chang}, \citenamefont {Yaji}, \citenamefont {Yuan}, \citenamefont {Zhang}, \citenamefont {Kuroda}, \citenamefont {Bian}, \citenamefont {Guo}, \citenamefont {Lu}, \citenamefont {Chang}, \citenamefont {Alidoust}, \citenamefont {Zheng}, \citenamefont {Lee}, \citenamefont {Huang}, \citenamefont {Hsu}, \citenamefont {Jeng}, \citenamefont {Bansil}, \citenamefont {Neupert}, \citenamefont {Komori}, \citenamefont {Kondo}, \citenamefont {Shin}, \citenamefont {Lin}, \citenamefont {Jia},\ and\ \citenamefont {Hasan}}]{arcd}%
  \BibitemOpen
  \bibfield  {author} {\bibinfo {author} {\bibfnamefont {S.-Y.}\ \bibnamefont {Xu}}, \bibinfo {author} {\bibfnamefont {I.}~\bibnamefont {Belopolski}}, \bibinfo {author} {\bibfnamefont {D.~S.}\ \bibnamefont {Sanchez}}, \bibinfo {author} {\bibfnamefont {M.}~\bibnamefont {Neupane}}, \bibinfo {author} {\bibfnamefont {G.}~\bibnamefont {Chang}}, \bibinfo {author} {\bibfnamefont {K.}~\bibnamefont {Yaji}}, \bibinfo {author} {\bibfnamefont {Z.}~\bibnamefont {Yuan}}, \bibinfo {author} {\bibfnamefont {C.}~\bibnamefont {Zhang}}, \bibinfo {author} {\bibfnamefont {K.}~\bibnamefont {Kuroda}}, \bibinfo {author} {\bibfnamefont {G.}~\bibnamefont {Bian}}, \bibinfo {author} {\bibfnamefont {C.}~\bibnamefont {Guo}}, \bibinfo {author} {\bibfnamefont {H.}~\bibnamefont {Lu}}, \bibinfo {author} {\bibfnamefont {T.-R.}\ \bibnamefont {Chang}}, \bibinfo {author} {\bibfnamefont {N.}~\bibnamefont {Alidoust}}, \bibinfo {author} {\bibfnamefont {H.}~\bibnamefont {Zheng}}, \bibinfo {author} {\bibfnamefont {C.-C.}\ \bibnamefont {Lee}}, \bibinfo
  {author} {\bibfnamefont {S.-M.}\ \bibnamefont {Huang}}, \bibinfo {author} {\bibfnamefont {C.-H.}\ \bibnamefont {Hsu}}, \bibinfo {author} {\bibfnamefont {H.-T.}\ \bibnamefont {Jeng}}, \bibinfo {author} {\bibfnamefont {A.}~\bibnamefont {Bansil}}, \bibinfo {author} {\bibfnamefont {T.}~\bibnamefont {Neupert}}, \bibinfo {author} {\bibfnamefont {F.}~\bibnamefont {Komori}}, \bibinfo {author} {\bibfnamefont {T.}~\bibnamefont {Kondo}}, \bibinfo {author} {\bibfnamefont {S.}~\bibnamefont {Shin}}, \bibinfo {author} {\bibfnamefont {H.}~\bibnamefont {Lin}}, \bibinfo {author} {\bibfnamefont {S.}~\bibnamefont {Jia}},\ and\ \bibinfo {author} {\bibfnamefont {M.~Z.}\ \bibnamefont {Hasan}},\ }\bibfield  {title} {\bibinfo {title} {Spin polarization and texture of the fermi arcs in the weyl fermion semimetal taas},\ }\href {https://doi.org/10.1103/PhysRevLett.116.096801} {\bibfield  {journal} {\bibinfo  {journal} {Phys. Rev. Lett.}\ }\textbf {\bibinfo {volume} {116}},\ \bibinfo {pages} {096801} (\bibinfo {year}
  {2016})}\BibitemShut {NoStop}%
\bibitem [{\citenamefont {Nielsen}\ and\ \citenamefont {Ninomiya}(1981{\natexlab{a}})}]{nielsen1981absence1}%
  \BibitemOpen
  \bibfield  {author} {\bibinfo {author} {\bibfnamefont {H.~B.}\ \bibnamefont {Nielsen}}\ and\ \bibinfo {author} {\bibfnamefont {M.}~\bibnamefont {Ninomiya}},\ }\bibfield  {title} {\bibinfo {title} {Absence of neutrinos on a lattice:(i). proof by homotopy theory},\ }\href {https://www.sciencedirect.com/science/article/pii/0550321381903618} {\bibfield  {journal} {\bibinfo  {journal} {Nuclear Physics B}\ }\textbf {\bibinfo {volume} {185}},\ \bibinfo {pages} {20} (\bibinfo {year} {1981}{\natexlab{a}})}\BibitemShut {NoStop}%
\bibitem [{\citenamefont {Nielsen}\ and\ \citenamefont {Ninomiya}(1981{\natexlab{b}})}]{nielsen1981absence2}%
  \BibitemOpen
  \bibfield  {author} {\bibinfo {author} {\bibfnamefont {H.~B.}\ \bibnamefont {Nielsen}}\ and\ \bibinfo {author} {\bibfnamefont {M.}~\bibnamefont {Ninomiya}},\ }\bibfield  {title} {\bibinfo {title} {Absence of neutrinos on a lattice:(ii). intuitive topological proof},\ }\href {https://www.sciencedirect.com/science/article/abs/pii/0550321381905241#:~:text=An%20intuitive%20topological%20proof%20is,preceding%20paper%20(Absence%20I).} {\bibfield  {journal} {\bibinfo  {journal} {Nuclear Physics B}\ }\textbf {\bibinfo {volume} {193}},\ \bibinfo {pages} {173} (\bibinfo {year} {1981}{\natexlab{b}})}\BibitemShut {NoStop}%
\bibitem [{\citenamefont {Lv}\ \emph {et~al.}(2015)\citenamefont {Lv}, \citenamefont {Weng}, \citenamefont {Fu}, \citenamefont {Wang}, \citenamefont {Miao}, \citenamefont {Ma}, \citenamefont {Richard}, \citenamefont {Huang}, \citenamefont {Zhao}, \citenamefont {Chen} \emph {et~al.}}]{lv2015experimental}%
  \BibitemOpen
  \bibfield  {author} {\bibinfo {author} {\bibfnamefont {B.}~\bibnamefont {Lv}}, \bibinfo {author} {\bibfnamefont {H.}~\bibnamefont {Weng}}, \bibinfo {author} {\bibfnamefont {B.}~\bibnamefont {Fu}}, \bibinfo {author} {\bibfnamefont {X.~P.}\ \bibnamefont {Wang}}, \bibinfo {author} {\bibfnamefont {H.}~\bibnamefont {Miao}}, \bibinfo {author} {\bibfnamefont {J.}~\bibnamefont {Ma}}, \bibinfo {author} {\bibfnamefont {P.}~\bibnamefont {Richard}}, \bibinfo {author} {\bibfnamefont {X.}~\bibnamefont {Huang}}, \bibinfo {author} {\bibfnamefont {L.}~\bibnamefont {Zhao}}, \bibinfo {author} {\bibfnamefont {G.}~\bibnamefont {Chen}}, \emph {et~al.},\ }\bibfield  {title} {\bibinfo {title} {Experimental discovery of weyl semimetal taas},\ }\href {https://journals.aps.org/prx/abstract/10.1103/PhysRevX.5.031013} {\bibfield  {journal} {\bibinfo  {journal} {Physical Review X}\ }\textbf {\bibinfo {volume} {5}},\ \bibinfo {pages} {031013} (\bibinfo {year} {2015})}\BibitemShut {NoStop}%
\bibitem [{\citenamefont {Xu}\ \emph {et~al.}(2015)\citenamefont {Xu}, \citenamefont {Belopolski}, \citenamefont {Sanchez}, \citenamefont {Zhang}, \citenamefont {Chang}, \citenamefont {Guo}, \citenamefont {Bian}, \citenamefont {Yuan}, \citenamefont {Lu}, \citenamefont {Chang} \emph {et~al.}}]{xu2015experimental}%
  \BibitemOpen
  \bibfield  {author} {\bibinfo {author} {\bibfnamefont {S.-Y.}\ \bibnamefont {Xu}}, \bibinfo {author} {\bibfnamefont {I.}~\bibnamefont {Belopolski}}, \bibinfo {author} {\bibfnamefont {D.~S.}\ \bibnamefont {Sanchez}}, \bibinfo {author} {\bibfnamefont {C.}~\bibnamefont {Zhang}}, \bibinfo {author} {\bibfnamefont {G.}~\bibnamefont {Chang}}, \bibinfo {author} {\bibfnamefont {C.}~\bibnamefont {Guo}}, \bibinfo {author} {\bibfnamefont {G.}~\bibnamefont {Bian}}, \bibinfo {author} {\bibfnamefont {Z.}~\bibnamefont {Yuan}}, \bibinfo {author} {\bibfnamefont {H.}~\bibnamefont {Lu}}, \bibinfo {author} {\bibfnamefont {T.-R.}\ \bibnamefont {Chang}}, \emph {et~al.},\ }\bibfield  {title} {\bibinfo {title} {Experimental discovery of a topological weyl semimetal state in tap},\ }\href {https://www.science.org/doi/10.1126/sciadv.1501092} {\bibfield  {journal} {\bibinfo  {journal} {Science advances}\ }\textbf {\bibinfo {volume} {1}},\ \bibinfo {pages} {e1501092} (\bibinfo {year} {2015})}\BibitemShut {NoStop}%
\bibitem [{\citenamefont {Deng}\ \emph {et~al.}(2016)\citenamefont {Deng}, \citenamefont {Wan}, \citenamefont {Deng}, \citenamefont {Zhang}, \citenamefont {Ding}, \citenamefont {Wang}, \citenamefont {Yan}, \citenamefont {Huang}, \citenamefont {Zhang}, \citenamefont {Xu} \emph {et~al.}}]{deng2016experimental}%
  \BibitemOpen
  \bibfield  {author} {\bibinfo {author} {\bibfnamefont {K.}~\bibnamefont {Deng}}, \bibinfo {author} {\bibfnamefont {G.}~\bibnamefont {Wan}}, \bibinfo {author} {\bibfnamefont {P.}~\bibnamefont {Deng}}, \bibinfo {author} {\bibfnamefont {K.}~\bibnamefont {Zhang}}, \bibinfo {author} {\bibfnamefont {S.}~\bibnamefont {Ding}}, \bibinfo {author} {\bibfnamefont {E.}~\bibnamefont {Wang}}, \bibinfo {author} {\bibfnamefont {M.}~\bibnamefont {Yan}}, \bibinfo {author} {\bibfnamefont {H.}~\bibnamefont {Huang}}, \bibinfo {author} {\bibfnamefont {H.}~\bibnamefont {Zhang}}, \bibinfo {author} {\bibfnamefont {Z.}~\bibnamefont {Xu}}, \emph {et~al.},\ }\bibfield  {title} {\bibinfo {title} {Experimental observation of topological fermi arcs in type-ii weyl semimetal mote2},\ }\href {https://www.nature.com/articles/nphys3871} {\bibfield  {journal} {\bibinfo  {journal} {Nature Physics}\ }\textbf {\bibinfo {volume} {12}},\ \bibinfo {pages} {1105} (\bibinfo {year} {2016})}\BibitemShut {NoStop}%
\bibitem [{\citenamefont {Huang}\ \emph {et~al.}(2015{\natexlab{b}})\citenamefont {Huang}, \citenamefont {Zhao}, \citenamefont {Long}, \citenamefont {Wang}, \citenamefont {Chen}, \citenamefont {Yang}, \citenamefont {Liang}, \citenamefont {Xue}, \citenamefont {Weng}, \citenamefont {Fang}, \citenamefont {Dai},\ and\ \citenamefont {Chen}}]{PhysRevX.5.031023}%
  \BibitemOpen
  \bibfield  {author} {\bibinfo {author} {\bibfnamefont {X.}~\bibnamefont {Huang}}, \bibinfo {author} {\bibfnamefont {L.}~\bibnamefont {Zhao}}, \bibinfo {author} {\bibfnamefont {Y.}~\bibnamefont {Long}}, \bibinfo {author} {\bibfnamefont {P.}~\bibnamefont {Wang}}, \bibinfo {author} {\bibfnamefont {D.}~\bibnamefont {Chen}}, \bibinfo {author} {\bibfnamefont {Z.}~\bibnamefont {Yang}}, \bibinfo {author} {\bibfnamefont {H.}~\bibnamefont {Liang}}, \bibinfo {author} {\bibfnamefont {M.}~\bibnamefont {Xue}}, \bibinfo {author} {\bibfnamefont {H.}~\bibnamefont {Weng}}, \bibinfo {author} {\bibfnamefont {Z.}~\bibnamefont {Fang}}, \bibinfo {author} {\bibfnamefont {X.}~\bibnamefont {Dai}},\ and\ \bibinfo {author} {\bibfnamefont {G.}~\bibnamefont {Chen}},\ }\bibfield  {title} {\bibinfo {title} {Observation of the chiral-anomaly-induced negative magnetoresistance in 3d weyl semimetal taas},\ }\href {https://doi.org/10.1103/PhysRevX.5.031023} {\bibfield  {journal} {\bibinfo  {journal} {Phys. Rev. X}\ }\textbf {\bibinfo {volume}
  {5}},\ \bibinfo {pages} {031023} (\bibinfo {year} {2015}{\natexlab{b}})}\BibitemShut {NoStop}%
\bibitem [{\citenamefont {Lu}\ \emph {et~al.}(2024)\citenamefont {Lu}, \citenamefont {Reddy}, \citenamefont {Jeon}, \citenamefont {Mazza}, \citenamefont {Brahlek}, \citenamefont {Wu}, \citenamefont {Yang}, \citenamefont {Cook}, \citenamefont {Conner}, \citenamefont {Zhang} \emph {et~al.}}]{lu2024}%
  \BibitemOpen
  \bibfield  {author} {\bibinfo {author} {\bibfnamefont {Q.}~\bibnamefont {Lu}}, \bibinfo {author} {\bibfnamefont {P.~S.}\ \bibnamefont {Reddy}}, \bibinfo {author} {\bibfnamefont {H.}~\bibnamefont {Jeon}}, \bibinfo {author} {\bibfnamefont {A.~R.}\ \bibnamefont {Mazza}}, \bibinfo {author} {\bibfnamefont {M.}~\bibnamefont {Brahlek}}, \bibinfo {author} {\bibfnamefont {W.}~\bibnamefont {Wu}}, \bibinfo {author} {\bibfnamefont {S.~A.}\ \bibnamefont {Yang}}, \bibinfo {author} {\bibfnamefont {J.}~\bibnamefont {Cook}}, \bibinfo {author} {\bibfnamefont {C.}~\bibnamefont {Conner}}, \bibinfo {author} {\bibfnamefont {X.}~\bibnamefont {Zhang}}, \emph {et~al.},\ }\bibfield  {title} {\bibinfo {title} {Realization of a two-dimensional weyl semimetal and topological fermi strings},\ }\href {https://www.nature.com/articles/s41467-024-50329-6} {\bibfield  {journal} {\bibinfo  {journal} {Nature communications}\ }\textbf {\bibinfo {volume} {15}},\ \bibinfo {pages} {6001} (\bibinfo {year} {2024})}\BibitemShut {NoStop}%
\bibitem [{\citenamefont {Ruderman}\ and\ \citenamefont {Kittel}(1954)}]{ruderman1954indirect}%
  \BibitemOpen
  \bibfield  {author} {\bibinfo {author} {\bibfnamefont {M.~A.}\ \bibnamefont {Ruderman}}\ and\ \bibinfo {author} {\bibfnamefont {C.}~\bibnamefont {Kittel}},\ }\bibfield  {title} {\bibinfo {title} {Indirect exchange coupling of nuclear magnetic moments by conduction electrons},\ }\href {https://journals.aps.org/pr/pdf/10.1103/PhysRev.96.99} {\bibfield  {journal} {\bibinfo  {journal} {Physical Review}\ }\textbf {\bibinfo {volume} {96}},\ \bibinfo {pages} {99} (\bibinfo {year} {1954})}\BibitemShut {NoStop}%
\bibitem [{\citenamefont {Kasuya}(1956)}]{kasuya1956theory}%
  \BibitemOpen
  \bibfield  {author} {\bibinfo {author} {\bibfnamefont {T.}~\bibnamefont {Kasuya}},\ }\bibfield  {title} {\bibinfo {title} {A theory of metallic ferro-and antiferromagnetism on zener's model},\ }\href {https://academic.oup.com/ptp/article/16/1/45/1861363} {\bibfield  {journal} {\bibinfo  {journal} {Progress of theoretical physics}\ }\textbf {\bibinfo {volume} {16}},\ \bibinfo {pages} {45} (\bibinfo {year} {1956})}\BibitemShut {NoStop}%
\bibitem [{\citenamefont {Yosida}(1957)}]{yosida1957magnetic}%
  \BibitemOpen
  \bibfield  {author} {\bibinfo {author} {\bibfnamefont {K.}~\bibnamefont {Yosida}},\ }\bibfield  {title} {\bibinfo {title} {Magnetic properties of cu-mn alloys},\ }\href {https://journals.aps.org/pr/abstract/10.1103/PhysRev.106.893} {\bibfield  {journal} {\bibinfo  {journal} {Physical Review}\ }\textbf {\bibinfo {volume} {106}},\ \bibinfo {pages} {893} (\bibinfo {year} {1957})}\BibitemShut {NoStop}%
\bibitem [{\citenamefont {Kossak}\ \emph {et~al.}(2023)\citenamefont {Kossak}, \citenamefont {Huang}, \citenamefont {Reddy}, \citenamefont {Wolf},\ and\ \citenamefont {Beach}}]{spintronic1}%
  \BibitemOpen
  \bibfield  {author} {\bibinfo {author} {\bibfnamefont {A.~E.}\ \bibnamefont {Kossak}}, \bibinfo {author} {\bibfnamefont {M.}~\bibnamefont {Huang}}, \bibinfo {author} {\bibfnamefont {P.}~\bibnamefont {Reddy}}, \bibinfo {author} {\bibfnamefont {D.}~\bibnamefont {Wolf}},\ and\ \bibinfo {author} {\bibfnamefont {G.~S.}\ \bibnamefont {Beach}},\ }\bibfield  {title} {\bibinfo {title} {Voltage control of magnetic order in rkky coupled multilayers},\ }\href {https://www.science.org/doi/10.1126/sciadv.add0548} {\bibfield  {journal} {\bibinfo  {journal} {Science Advances}\ }\textbf {\bibinfo {volume} {9}},\ \bibinfo {pages} {eadd0548} (\bibinfo {year} {2023})}\BibitemShut {NoStop}%
\bibitem [{\citenamefont {Yarmohammadi}\ \emph {et~al.}(2023)\citenamefont {Yarmohammadi}, \citenamefont {Bukov},\ and\ \citenamefont {Kolodrubetz}}]{spintronic2}%
  \BibitemOpen
  \bibfield  {author} {\bibinfo {author} {\bibfnamefont {M.}~\bibnamefont {Yarmohammadi}}, \bibinfo {author} {\bibfnamefont {M.}~\bibnamefont {Bukov}},\ and\ \bibinfo {author} {\bibfnamefont {M.~H.}\ \bibnamefont {Kolodrubetz}},\ }\bibfield  {title} {\bibinfo {title} {Noncollinear twisted rkky interaction on the optically driven snte(001) surface},\ }\href {https://doi.org/10.1103/PhysRevB.107.054439} {\bibfield  {journal} {\bibinfo  {journal} {Phys. Rev. B}\ }\textbf {\bibinfo {volume} {107}},\ \bibinfo {pages} {054439} (\bibinfo {year} {2023})}\BibitemShut {NoStop}%
\bibitem [{\citenamefont {Gorman}\ \emph {et~al.}(2014)\citenamefont {Gorman}, \citenamefont {Duffy}, \citenamefont {Power},\ and\ \citenamefont {Ferreira}}]{spintronic3}%
  \BibitemOpen
  \bibfield  {author} {\bibinfo {author} {\bibfnamefont {P.~D.}\ \bibnamefont {Gorman}}, \bibinfo {author} {\bibfnamefont {J.~M.}\ \bibnamefont {Duffy}}, \bibinfo {author} {\bibfnamefont {S.~R.}\ \bibnamefont {Power}},\ and\ \bibinfo {author} {\bibfnamefont {M.~S.}\ \bibnamefont {Ferreira}},\ }\bibfield  {title} {\bibinfo {title} {Rkky interaction between extended magnetic defect lines in graphene},\ }\href {https://doi.org/10.1103/PhysRevB.90.125411} {\bibfield  {journal} {\bibinfo  {journal} {Phys. Rev. B}\ }\textbf {\bibinfo {volume} {90}},\ \bibinfo {pages} {125411} (\bibinfo {year} {2014})}\BibitemShut {NoStop}%
\bibitem [{\citenamefont {Hermenau}\ \emph {et~al.}(2019)\citenamefont {Hermenau}, \citenamefont {Brinker}, \citenamefont {Marciani}, \citenamefont {Steinbrecher}, \citenamefont {dos Santos~Dias}, \citenamefont {Wiesendanger}, \citenamefont {Lounis},\ and\ \citenamefont {Wiebe}}]{hermenau2019stabilizing}%
  \BibitemOpen
  \bibfield  {author} {\bibinfo {author} {\bibfnamefont {J.}~\bibnamefont {Hermenau}}, \bibinfo {author} {\bibfnamefont {S.}~\bibnamefont {Brinker}}, \bibinfo {author} {\bibfnamefont {M.}~\bibnamefont {Marciani}}, \bibinfo {author} {\bibfnamefont {M.}~\bibnamefont {Steinbrecher}}, \bibinfo {author} {\bibfnamefont {M.}~\bibnamefont {dos Santos~Dias}}, \bibinfo {author} {\bibfnamefont {R.}~\bibnamefont {Wiesendanger}}, \bibinfo {author} {\bibfnamefont {S.}~\bibnamefont {Lounis}},\ and\ \bibinfo {author} {\bibfnamefont {J.}~\bibnamefont {Wiebe}},\ }\bibfield  {title} {\bibinfo {title} {Stabilizing spin systems via symmetrically tailored rkky interactions},\ }\href {https://www.nature.com/articles/s41467-019-10516-2} {\bibfield  {journal} {\bibinfo  {journal} {Nature Communications}\ }\textbf {\bibinfo {volume} {10}},\ \bibinfo {pages} {2565} (\bibinfo {year} {2019})}\BibitemShut {NoStop}%
\bibitem [{\citenamefont {Imamura}\ \emph {et~al.}(2004)\citenamefont {Imamura}, \citenamefont {Bruno},\ and\ \citenamefont {Utsumi}}]{imamura2004twisted}%
  \BibitemOpen
  \bibfield  {author} {\bibinfo {author} {\bibfnamefont {H.}~\bibnamefont {Imamura}}, \bibinfo {author} {\bibfnamefont {P.}~\bibnamefont {Bruno}},\ and\ \bibinfo {author} {\bibfnamefont {Y.}~\bibnamefont {Utsumi}},\ }\bibfield  {title} {\bibinfo {title} {Twisted exchange interaction between localized spins embedded in a one-or two-dimensional electron gas with rashba spin-orbit coupling},\ }\href {https://journals.aps.org/prb/abstract/10.1103/PhysRevB.69.121303} {\bibfield  {journal} {\bibinfo  {journal} {Physical Review B}\ }\textbf {\bibinfo {volume} {69}},\ \bibinfo {pages} {121303} (\bibinfo {year} {2004})}\BibitemShut {NoStop}%
\bibitem [{\citenamefont {Schulz}\ \emph {et~al.}(2009)\citenamefont {Schulz}, \citenamefont {De~Martino}, \citenamefont {Ingenhoven},\ and\ \citenamefont {Egger}}]{schulz2009low}%
  \BibitemOpen
  \bibfield  {author} {\bibinfo {author} {\bibfnamefont {A.}~\bibnamefont {Schulz}}, \bibinfo {author} {\bibfnamefont {A.}~\bibnamefont {De~Martino}}, \bibinfo {author} {\bibfnamefont {P.}~\bibnamefont {Ingenhoven}},\ and\ \bibinfo {author} {\bibfnamefont {R.}~\bibnamefont {Egger}},\ }\bibfield  {title} {\bibinfo {title} {Low-energy theory and rkky interaction for interacting quantum wires with rashba spin-orbit coupling},\ }\href {https://journals.aps.org/prb/abstract/10.1103/PhysRevB.79.205432} {\bibfield  {journal} {\bibinfo  {journal} {Physical Review B}\ }\textbf {\bibinfo {volume} {79}},\ \bibinfo {pages} {205432} (\bibinfo {year} {2009})}\BibitemShut {NoStop}%
\bibitem [{\citenamefont {Gao}\ \emph {et~al.}(2009)\citenamefont {Gao}, \citenamefont {Chen}, \citenamefont {Xie},\ and\ \citenamefont {Zhang}}]{gao2009plane}%
  \BibitemOpen
  \bibfield  {author} {\bibinfo {author} {\bibfnamefont {J.}~\bibnamefont {Gao}}, \bibinfo {author} {\bibfnamefont {W.}~\bibnamefont {Chen}}, \bibinfo {author} {\bibfnamefont {X.}~\bibnamefont {Xie}},\ and\ \bibinfo {author} {\bibfnamefont {F.-c.}\ \bibnamefont {Zhang}},\ }\bibfield  {title} {\bibinfo {title} {In-plane noncollinear exchange coupling mediated by helical edge states in quantum spin hall systems},\ }\href {https://journals.aps.org/prb/abstract/10.1103/PhysRevB.80.241302} {\bibfield  {journal} {\bibinfo  {journal} {Physical Review B}\ }\textbf {\bibinfo {volume} {80}},\ \bibinfo {pages} {241302} (\bibinfo {year} {2009})}\BibitemShut {NoStop}%
\bibitem [{\citenamefont {Liu}\ \emph {et~al.}(2009)\citenamefont {Liu}, \citenamefont {Liu}, \citenamefont {Xu}, \citenamefont {Qi},\ and\ \citenamefont {Zhang}}]{liu2009magnetic}%
  \BibitemOpen
  \bibfield  {author} {\bibinfo {author} {\bibfnamefont {Q.}~\bibnamefont {Liu}}, \bibinfo {author} {\bibfnamefont {C.-X.}\ \bibnamefont {Liu}}, \bibinfo {author} {\bibfnamefont {C.}~\bibnamefont {Xu}}, \bibinfo {author} {\bibfnamefont {X.-L.}\ \bibnamefont {Qi}},\ and\ \bibinfo {author} {\bibfnamefont {S.-C.}\ \bibnamefont {Zhang}},\ }\bibfield  {title} {\bibinfo {title} {Magnetic impurities on the surface of a topological insulator},\ }\href {https://journals.aps.org/prl/abstract/10.1103/PhysRevLett.102.156603} {\bibfield  {journal} {\bibinfo  {journal} {Physical review letters}\ }\textbf {\bibinfo {volume} {102}},\ \bibinfo {pages} {156603} (\bibinfo {year} {2009})}\BibitemShut {NoStop}%
\bibitem [{\citenamefont {Sherafati}\ and\ \citenamefont {Satpathy}(2011)}]{sherafati2011rkky}%
  \BibitemOpen
  \bibfield  {author} {\bibinfo {author} {\bibfnamefont {M.}~\bibnamefont {Sherafati}}\ and\ \bibinfo {author} {\bibfnamefont {S.}~\bibnamefont {Satpathy}},\ }\bibfield  {title} {\bibinfo {title} {Rkky interaction in graphene from the lattice green’s function},\ }\href {https://link.aps.org/doi/10.1103/PhysRevB.83.165425} {\bibfield  {journal} {\bibinfo  {journal} {Physical Review B}\ }\textbf {\bibinfo {volume} {83}},\ \bibinfo {pages} {165425} (\bibinfo {year} {2011})}\BibitemShut {NoStop}%
\bibitem [{\citenamefont {Kogan}(2011)}]{kogan2011rkky}%
  \BibitemOpen
  \bibfield  {author} {\bibinfo {author} {\bibfnamefont {E.}~\bibnamefont {Kogan}},\ }\bibfield  {title} {\bibinfo {title} {Rkky interaction in graphene},\ }\href {https://journals.aps.org/prb/abstract/10.1103/PhysRevB.84.115119} {\bibfield  {journal} {\bibinfo  {journal} {Physical Review B}\ }\textbf {\bibinfo {volume} {84}},\ \bibinfo {pages} {115119} (\bibinfo {year} {2011})}\BibitemShut {NoStop}%
\bibitem [{\citenamefont {Mastrogiuseppe}\ \emph {et~al.}(2016)\citenamefont {Mastrogiuseppe}, \citenamefont {Sandler},\ and\ \citenamefont {Ulloa}}]{mastrogiuseppe2016hybridization}%
  \BibitemOpen
  \bibfield  {author} {\bibinfo {author} {\bibfnamefont {D.}~\bibnamefont {Mastrogiuseppe}}, \bibinfo {author} {\bibfnamefont {N.}~\bibnamefont {Sandler}},\ and\ \bibinfo {author} {\bibfnamefont {S.}~\bibnamefont {Ulloa}},\ }\bibfield  {title} {\bibinfo {title} {Hybridization and anisotropy in the exchange interaction in three-dimensional dirac semimetals},\ }\href {https://journals.aps.org/prb/abstract/10.1103/PhysRevB.93.094433} {\bibfield  {journal} {\bibinfo  {journal} {Physical Review B}\ }\textbf {\bibinfo {volume} {93}},\ \bibinfo {pages} {094433} (\bibinfo {year} {2016})}\BibitemShut {NoStop}%
\bibitem [{\citenamefont {Zare}\ \emph {et~al.}(2016)\citenamefont {Zare}, \citenamefont {Parhizgar},\ and\ \citenamefont {Asgari}}]{zare2016topological}%
  \BibitemOpen
  \bibfield  {author} {\bibinfo {author} {\bibfnamefont {M.}~\bibnamefont {Zare}}, \bibinfo {author} {\bibfnamefont {F.}~\bibnamefont {Parhizgar}},\ and\ \bibinfo {author} {\bibfnamefont {R.}~\bibnamefont {Asgari}},\ }\bibfield  {title} {\bibinfo {title} {Topological phase and edge states dependence of the rkky interaction in zigzag silicene nanoribbon},\ }\href {https://journals.aps.org/prb/abstract/10.1103/PhysRevB.94.045443} {\bibfield  {journal} {\bibinfo  {journal} {Physical Review B}\ }\textbf {\bibinfo {volume} {94}},\ \bibinfo {pages} {045443} (\bibinfo {year} {2016})}\BibitemShut {NoStop}%
\bibitem [{\citenamefont {Yang}\ \emph {et~al.}(2016)\citenamefont {Yang}, \citenamefont {Hsu}, \citenamefont {Stano}, \citenamefont {Klinovaja},\ and\ \citenamefont {Loss}}]{yang2016long}%
  \BibitemOpen
  \bibfield  {author} {\bibinfo {author} {\bibfnamefont {G.}~\bibnamefont {Yang}}, \bibinfo {author} {\bibfnamefont {C.-H.}\ \bibnamefont {Hsu}}, \bibinfo {author} {\bibfnamefont {P.}~\bibnamefont {Stano}}, \bibinfo {author} {\bibfnamefont {J.}~\bibnamefont {Klinovaja}},\ and\ \bibinfo {author} {\bibfnamefont {D.}~\bibnamefont {Loss}},\ }\bibfield  {title} {\bibinfo {title} {Long-distance entanglement of spin qubits via quantum hall edge states},\ }\href {https://journals.aps.org/prb/abstract/10.1103/PhysRevB.93.075301} {\bibfield  {journal} {\bibinfo  {journal} {Physical Review B}\ }\textbf {\bibinfo {volume} {93}},\ \bibinfo {pages} {075301} (\bibinfo {year} {2016})}\BibitemShut {NoStop}%
\bibitem [{\citenamefont {Hsu}\ \emph {et~al.}(2017)\citenamefont {Hsu}, \citenamefont {Stano}, \citenamefont {Klinovaja},\ and\ \citenamefont {Loss}}]{hsu2017nuclear}%
  \BibitemOpen
  \bibfield  {author} {\bibinfo {author} {\bibfnamefont {C.-H.}\ \bibnamefont {Hsu}}, \bibinfo {author} {\bibfnamefont {P.}~\bibnamefont {Stano}}, \bibinfo {author} {\bibfnamefont {J.}~\bibnamefont {Klinovaja}},\ and\ \bibinfo {author} {\bibfnamefont {D.}~\bibnamefont {Loss}},\ }\bibfield  {title} {\bibinfo {title} {Nuclear-spin-induced localization of edge states in two-dimensional topological insulators},\ }\href {https://journals.aps.org/prb/abstract/10.1103/PhysRevB.96.081405} {\bibfield  {journal} {\bibinfo  {journal} {Physical Review B}\ }\textbf {\bibinfo {volume} {96}},\ \bibinfo {pages} {081405} (\bibinfo {year} {2017})}\BibitemShut {NoStop}%
\bibitem [{\citenamefont {Hsu}\ \emph {et~al.}(2018)\citenamefont {Hsu}, \citenamefont {Stano}, \citenamefont {Klinovaja},\ and\ \citenamefont {Loss}}]{hsu2018effects}%
  \BibitemOpen
  \bibfield  {author} {\bibinfo {author} {\bibfnamefont {C.-H.}\ \bibnamefont {Hsu}}, \bibinfo {author} {\bibfnamefont {P.}~\bibnamefont {Stano}}, \bibinfo {author} {\bibfnamefont {J.}~\bibnamefont {Klinovaja}},\ and\ \bibinfo {author} {\bibfnamefont {D.}~\bibnamefont {Loss}},\ }\bibfield  {title} {\bibinfo {title} {Effects of nuclear spins on the transport properties of the edge of two-dimensional topological insulators},\ }\href {https://link.aps.org/doi/10.1103/PhysRevB.97.125432} {\bibfield  {journal} {\bibinfo  {journal} {Physical Review B}\ }\textbf {\bibinfo {volume} {97}},\ \bibinfo {pages} {125432} (\bibinfo {year} {2018})}\BibitemShut {NoStop}%
\bibitem [{\citenamefont {Lee}\ and\ \citenamefont {Lee}(2015)}]{lee2015electrical}%
  \BibitemOpen
  \bibfield  {author} {\bibinfo {author} {\bibfnamefont {Y.-W.}\ \bibnamefont {Lee}}\ and\ \bibinfo {author} {\bibfnamefont {Y.-L.}\ \bibnamefont {Lee}},\ }\bibfield  {title} {\bibinfo {title} {Electrical control and interaction effects of the rkky interaction in helical liquids},\ }\href {https://journals.aps.org/prb/abstract/10.1103/PhysRevB.91.214431} {\bibfield  {journal} {\bibinfo  {journal} {Physical Review B}\ }\textbf {\bibinfo {volume} {91}},\ \bibinfo {pages} {214431} (\bibinfo {year} {2015})}\BibitemShut {NoStop}%
\bibitem [{\citenamefont {Reja}\ \emph {et~al.}(2017)\citenamefont {Reja}, \citenamefont {Fertig}, \citenamefont {Brey},\ and\ \citenamefont {Zhang}}]{reja2017surface}%
  \BibitemOpen
  \bibfield  {author} {\bibinfo {author} {\bibfnamefont {S.}~\bibnamefont {Reja}}, \bibinfo {author} {\bibfnamefont {H.}~\bibnamefont {Fertig}}, \bibinfo {author} {\bibfnamefont {L.}~\bibnamefont {Brey}},\ and\ \bibinfo {author} {\bibfnamefont {S.}~\bibnamefont {Zhang}},\ }\bibfield  {title} {\bibinfo {title} {Surface magnetism in topological crystalline insulators},\ }\href {https://journals.aps.org/prb/abstract/10.1103/PhysRevB.96.201111} {\bibfield  {journal} {\bibinfo  {journal} {Physical Review B}\ }\textbf {\bibinfo {volume} {96}},\ \bibinfo {pages} {201111} (\bibinfo {year} {2017})}\BibitemShut {NoStop}%
\bibitem [{\citenamefont {Reja}\ \emph {et~al.}(2019)\citenamefont {Reja}, \citenamefont {Fertig},\ and\ \citenamefont {Brey}}]{reja2019spin}%
  \BibitemOpen
  \bibfield  {author} {\bibinfo {author} {\bibfnamefont {S.}~\bibnamefont {Reja}}, \bibinfo {author} {\bibfnamefont {H.~A.}\ \bibnamefont {Fertig}},\ and\ \bibinfo {author} {\bibfnamefont {L.}~\bibnamefont {Brey}},\ }\bibfield  {title} {\bibinfo {title} {Spin stiffness and domain walls in dirac-electron mediated magnets},\ }\href {https://journals.aps.org/prb/abstract/10.1103/PhysRevB.99.045427} {\bibfield  {journal} {\bibinfo  {journal} {Physical Review B}\ }\textbf {\bibinfo {volume} {99}},\ \bibinfo {pages} {045427} (\bibinfo {year} {2019})}\BibitemShut {NoStop}%
\bibitem [{\citenamefont {Verma}\ and\ \citenamefont {Kundu}(2019{\natexlab{a}})}]{verma2019nonlocal}%
  \BibitemOpen
  \bibfield  {author} {\bibinfo {author} {\bibfnamefont {S.}~\bibnamefont {Verma}}\ and\ \bibinfo {author} {\bibfnamefont {A.}~\bibnamefont {Kundu}},\ }\bibfield  {title} {\bibinfo {title} {Nonlocal control of spin-spin correlations in a finite-geometry helical edge},\ }\href {https://journals.aps.org/prb/abstract/10.1103/PhysRevB.99.121409} {\bibfield  {journal} {\bibinfo  {journal} {Physical Review B}\ }\textbf {\bibinfo {volume} {99}},\ \bibinfo {pages} {121409} (\bibinfo {year} {2019}{\natexlab{a}})}\BibitemShut {NoStop}%
\bibitem [{\citenamefont {Chang}\ \emph {et~al.}(2015)\citenamefont {Chang}, \citenamefont {Zhou}, \citenamefont {Wang}, \citenamefont {Shan},\ and\ \citenamefont {Xiao}}]{chang2015rkky}%
  \BibitemOpen
  \bibfield  {author} {\bibinfo {author} {\bibfnamefont {H.-R.}\ \bibnamefont {Chang}}, \bibinfo {author} {\bibfnamefont {J.}~\bibnamefont {Zhou}}, \bibinfo {author} {\bibfnamefont {S.-X.}\ \bibnamefont {Wang}}, \bibinfo {author} {\bibfnamefont {W.-Y.}\ \bibnamefont {Shan}},\ and\ \bibinfo {author} {\bibfnamefont {D.}~\bibnamefont {Xiao}},\ }\bibfield  {title} {\bibinfo {title} {Rkky interaction of magnetic impurities in dirac and weyl semimetals},\ }\href {https://journals.aps.org/prb/abstract/10.1103/PhysRevB.92.241103} {\bibfield  {journal} {\bibinfo  {journal} {Physical Review B}\ }\textbf {\bibinfo {volume} {92}},\ \bibinfo {pages} {241103} (\bibinfo {year} {2015})}\BibitemShut {NoStop}%
\bibitem [{\citenamefont {Sun}\ and\ \citenamefont {Wang}(2017)}]{sun2017rkky}%
  \BibitemOpen
  \bibfield  {author} {\bibinfo {author} {\bibfnamefont {Y.}~\bibnamefont {Sun}}\ and\ \bibinfo {author} {\bibfnamefont {A.}~\bibnamefont {Wang}},\ }\bibfield  {title} {\bibinfo {title} {Rkky interaction of magnetic impurities in multi-weyl semimetals},\ }\href {https://iopscience.iop.org/article/10.1088/1361-648X/aa8932} {\bibfield  {journal} {\bibinfo  {journal} {Journal of Physics: Condensed Matter}\ }\textbf {\bibinfo {volume} {29}},\ \bibinfo {pages} {435306} (\bibinfo {year} {2017})}\BibitemShut {NoStop}%
\bibitem [{\citenamefont {Hosseini}\ and\ \citenamefont {Askari}(2015)}]{hosseini2015ruderman}%
  \BibitemOpen
  \bibfield  {author} {\bibinfo {author} {\bibfnamefont {M.~V.}\ \bibnamefont {Hosseini}}\ and\ \bibinfo {author} {\bibfnamefont {M.}~\bibnamefont {Askari}},\ }\bibfield  {title} {\bibinfo {title} {Ruderman-kittel-kasuya-yosida interaction in weyl semimetals},\ }\href {https://journals.aps.org/prb/pdf/10.1103/PhysRevB.92.224435} {\bibfield  {journal} {\bibinfo  {journal} {Physical Review B}\ }\textbf {\bibinfo {volume} {92}},\ \bibinfo {pages} {224435} (\bibinfo {year} {2015})}\BibitemShut {NoStop}%
\bibitem [{\citenamefont {Duan}\ \emph {et~al.}(2018)\citenamefont {Duan}, \citenamefont {Zheng}, \citenamefont {Fu}, \citenamefont {Wang}, \citenamefont {Liu}, \citenamefont {Wang},\ and\ \citenamefont {Yang}}]{duan2018indirect}%
  \BibitemOpen
  \bibfield  {author} {\bibinfo {author} {\bibfnamefont {H.-J.}\ \bibnamefont {Duan}}, \bibinfo {author} {\bibfnamefont {S.-H.}\ \bibnamefont {Zheng}}, \bibinfo {author} {\bibfnamefont {P.-H.}\ \bibnamefont {Fu}}, \bibinfo {author} {\bibfnamefont {R.-Q.}\ \bibnamefont {Wang}}, \bibinfo {author} {\bibfnamefont {J.-F.}\ \bibnamefont {Liu}}, \bibinfo {author} {\bibfnamefont {G.-H.}\ \bibnamefont {Wang}},\ and\ \bibinfo {author} {\bibfnamefont {M.}~\bibnamefont {Yang}},\ }\bibfield  {title} {\bibinfo {title} {Indirect magnetic interaction mediated by fermi arc and boundary reflection near weyl semimetal surface},\ }\href {https://iopscience.iop.org/article/10.1088/1367-2630/aae3f9} {\bibfield  {journal} {\bibinfo  {journal} {New Journal of Physics}\ }\textbf {\bibinfo {volume} {20}},\ \bibinfo {pages} {103008} (\bibinfo {year} {2018})}\BibitemShut {NoStop}%
\bibitem [{\citenamefont {Ma}\ \emph {et~al.}(2018)\citenamefont {Ma}, \citenamefont {Chen}, \citenamefont {Liu},\ and\ \citenamefont {Xie}}]{ma2018kondo}%
  \BibitemOpen
  \bibfield  {author} {\bibinfo {author} {\bibfnamefont {D.}~\bibnamefont {Ma}}, \bibinfo {author} {\bibfnamefont {H.}~\bibnamefont {Chen}}, \bibinfo {author} {\bibfnamefont {H.}~\bibnamefont {Liu}},\ and\ \bibinfo {author} {\bibfnamefont {X.}~\bibnamefont {Xie}},\ }\bibfield  {title} {\bibinfo {title} {Kondo effect with weyl semimetal fermi arcs},\ }\href {https://journals.aps.org/prb/abstract/10.1103/PhysRevB.97.045148} {\bibfield  {journal} {\bibinfo  {journal} {Physical Review B}\ }\textbf {\bibinfo {volume} {97}},\ \bibinfo {pages} {045148} (\bibinfo {year} {2018})}\BibitemShut {NoStop}%
\bibitem [{\citenamefont {Verma}\ \emph {et~al.}(2020)\citenamefont {Verma}, \citenamefont {Giri}, \citenamefont {Fertig},\ and\ \citenamefont {Kundu}}]{verma2020rkky}%
  \BibitemOpen
  \bibfield  {author} {\bibinfo {author} {\bibfnamefont {S.}~\bibnamefont {Verma}}, \bibinfo {author} {\bibfnamefont {D.}~\bibnamefont {Giri}}, \bibinfo {author} {\bibfnamefont {H.~A.}\ \bibnamefont {Fertig}},\ and\ \bibinfo {author} {\bibfnamefont {A.}~\bibnamefont {Kundu}},\ }\bibfield  {title} {\bibinfo {title} {Rkky coupling in weyl semimetal thin films},\ }\href {https://link.aps.org/doi/10.1103/PhysRevB.101.085419} {\bibfield  {journal} {\bibinfo  {journal} {Physical Review B}\ }\textbf {\bibinfo {volume} {101}},\ \bibinfo {pages} {085419} (\bibinfo {year} {2020})}\BibitemShut {NoStop}%
\bibitem [{\citenamefont {Klinovaja}\ and\ \citenamefont {Loss}(2013)}]{graphenerkky1}%
  \BibitemOpen
  \bibfield  {author} {\bibinfo {author} {\bibfnamefont {J.}~\bibnamefont {Klinovaja}}\ and\ \bibinfo {author} {\bibfnamefont {D.}~\bibnamefont {Loss}},\ }\bibfield  {title} {\bibinfo {title} {Rkky interaction in carbon nanotubes and graphene nanoribbons},\ }\href {https://journals.aps.org/prb/abstract/10.1103/PhysRevB.87.045422} {\bibfield  {journal} {\bibinfo  {journal} {Phys. Rev. B}\ }\textbf {\bibinfo {volume} {87}},\ \bibinfo {pages} {045422} (\bibinfo {year} {2013})}\BibitemShut {NoStop}%
\bibitem [{\citenamefont {Nguyen}\ and\ \citenamefont {Chu}(2020)}]{graphenerkky2}%
  \BibitemOpen
  \bibfield  {author} {\bibinfo {author} {\bibfnamefont {V.~M.}\ \bibnamefont {Nguyen}}\ and\ \bibinfo {author} {\bibfnamefont {C.~S.}\ \bibnamefont {Chu}},\ }\bibfield  {title} {\bibinfo {title} {Large rkky coupling from multiple scattering in armchair graphene nanoribbons},\ }\href {https://journals.aps.org/prb/abstract/10.1103/PhysRevB.101.195419} {\bibfield  {journal} {\bibinfo  {journal} {Phys. Rev. B}\ }\textbf {\bibinfo {volume} {101}},\ \bibinfo {pages} {195419} (\bibinfo {year} {2020})}\BibitemShut {NoStop}%
\bibitem [{\citenamefont {Igarashi}\ and\ \citenamefont {Koshino}(2017)}]{igarashi2017magnetotransport}%
  \BibitemOpen
  \bibfield  {author} {\bibinfo {author} {\bibfnamefont {A.}~\bibnamefont {Igarashi}}\ and\ \bibinfo {author} {\bibfnamefont {M.}~\bibnamefont {Koshino}},\ }\bibfield  {title} {\bibinfo {title} {Magnetotransport in weyl semimetal nanowires},\ }\href {https://journals.aps.org/prb/abstract/10.1103/PhysRevB.95.195306} {\bibfield  {journal} {\bibinfo  {journal} {Physical Review B}\ }\textbf {\bibinfo {volume} {95}},\ \bibinfo {pages} {195306} (\bibinfo {year} {2017})}\BibitemShut {NoStop}%
\bibitem [{\citenamefont {Erementchouk}\ and\ \citenamefont {Mazumder}(2018)}]{PhysRevB.97.035429}%
  \BibitemOpen
  \bibfield  {author} {\bibinfo {author} {\bibfnamefont {M.}~\bibnamefont {Erementchouk}}\ and\ \bibinfo {author} {\bibfnamefont {P.}~\bibnamefont {Mazumder}},\ }\bibfield  {title} {\bibinfo {title} {Weyl fermions in cylindrical wires},\ }\href {https://doi.org/10.1103/PhysRevB.97.035429} {\bibfield  {journal} {\bibinfo  {journal} {Phys. Rev. B}\ }\textbf {\bibinfo {volume} {97}},\ \bibinfo {pages} {035429} (\bibinfo {year} {2018})}\BibitemShut {NoStop}%
\bibitem [{\citenamefont {Kaladzhyan}\ and\ \citenamefont {Bardarson}(2019)}]{kaladzhyan2019quantized}%
  \BibitemOpen
  \bibfield  {author} {\bibinfo {author} {\bibfnamefont {V.}~\bibnamefont {Kaladzhyan}}\ and\ \bibinfo {author} {\bibfnamefont {J.~H.}\ \bibnamefont {Bardarson}},\ }\bibfield  {title} {\bibinfo {title} {Quantized fermi arc mediated transport in weyl semimetal nanowires},\ }\href {https://link.aps.org/doi/10.1103/PhysRevB.100.085424} {\bibfield  {journal} {\bibinfo  {journal} {Physical Review B}\ }\textbf {\bibinfo {volume} {100}},\ \bibinfo {pages} {085424} (\bibinfo {year} {2019})}\BibitemShut {NoStop}%
\bibitem [{\citenamefont {De~Martino}\ \emph {et~al.}(2021)\citenamefont {De~Martino}, \citenamefont {Dorn}, \citenamefont {Buccheri},\ and\ \citenamefont {Egger}}]{PhysRevB.104.155425}%
  \BibitemOpen
  \bibfield  {author} {\bibinfo {author} {\bibfnamefont {A.}~\bibnamefont {De~Martino}}, \bibinfo {author} {\bibfnamefont {K.}~\bibnamefont {Dorn}}, \bibinfo {author} {\bibfnamefont {F.}~\bibnamefont {Buccheri}},\ and\ \bibinfo {author} {\bibfnamefont {R.}~\bibnamefont {Egger}},\ }\bibfield  {title} {\bibinfo {title} {Phonon-induced magnetoresistivity of weyl semimetal nanowires},\ }\href {https://doi.org/10.1103/PhysRevB.104.155425} {\bibfield  {journal} {\bibinfo  {journal} {Phys. Rev. B}\ }\textbf {\bibinfo {volume} {104}},\ \bibinfo {pages} {155425} (\bibinfo {year} {2021})}\BibitemShut {NoStop}%
\bibitem [{\citenamefont {Jungwirth}\ \emph {et~al.}(2006)\citenamefont {Jungwirth}, \citenamefont {Sinova}, \citenamefont {Ma\ifmmode~\check{s}\else \v{s}\fi{}ek}, \citenamefont {Ku\ifmmode~\check{c}\else \v{c}\fi{}era},\ and\ \citenamefont {MacDonald}}]{RevModPhys.78.809}%
  \BibitemOpen
  \bibfield  {author} {\bibinfo {author} {\bibfnamefont {T.}~\bibnamefont {Jungwirth}}, \bibinfo {author} {\bibfnamefont {J.}~\bibnamefont {Sinova}}, \bibinfo {author} {\bibfnamefont {J.}~\bibnamefont {Ma\ifmmode~\check{s}\else \v{s}\fi{}ek}}, \bibinfo {author} {\bibfnamefont {J.}~\bibnamefont {Ku\ifmmode~\check{c}\else \v{c}\fi{}era}},\ and\ \bibinfo {author} {\bibfnamefont {A.~H.}\ \bibnamefont {MacDonald}},\ }\bibfield  {title} {\bibinfo {title} {Theory of ferromagnetic (iii,mn)v semiconductors},\ }\href {https://journals.aps.org/rmp/abstract/10.1103/RevModPhys.78.809} {\bibfield  {journal} {\bibinfo  {journal} {Rev. Mod. Phys.}\ }\textbf {\bibinfo {volume} {78}},\ \bibinfo {pages} {809} (\bibinfo {year} {2006})}\BibitemShut {NoStop}%
\bibitem [{\citenamefont {Coleman}(2017)}]{coleman2017introduction}%
  \BibitemOpen
  \bibfield  {author} {\bibinfo {author} {\bibfnamefont {P.}~\bibnamefont {Coleman}},\ }\href {https://www.cambridge.org/core/books/introduction-to-manybody-physics/B7598FC1FCEE0285F5EC767E835854C8} {\emph {\bibinfo {title} {Introduction to many-body physics}}},\ Vol.~\bibinfo {volume} {1}\ (\bibinfo  {publisher} {Cambridge University Press},\ \bibinfo {year} {2017})\BibitemShut {NoStop}%
\bibitem [{\citenamefont {Verma}\ and\ \citenamefont {Kundu}(2019{\natexlab{b}})}]{sonurkkyqsh}%
  \BibitemOpen
  \bibfield  {author} {\bibinfo {author} {\bibfnamefont {S.}~\bibnamefont {Verma}}\ and\ \bibinfo {author} {\bibfnamefont {A.}~\bibnamefont {Kundu}},\ }\bibfield  {title} {\bibinfo {title} {Nonlocal control of spin-spin correlations in a finite-geometry helical edge},\ }\href {https://journals.aps.org/prb/abstract/10.1103/PhysRevB.99.121409} {\bibfield  {journal} {\bibinfo  {journal} {Phys. Rev. B}\ }\textbf {\bibinfo {volume} {99}},\ \bibinfo {pages} {121409} (\bibinfo {year} {2019}{\natexlab{b}})}\BibitemShut {NoStop}%
\bibitem [{\citenamefont {Mugiraneza}\ and\ \citenamefont {Hallas}(2022)}]{mugiraneza2022tutorial}%
  \BibitemOpen
  \bibfield  {author} {\bibinfo {author} {\bibfnamefont {S.}~\bibnamefont {Mugiraneza}}\ and\ \bibinfo {author} {\bibfnamefont {A.~M.}\ \bibnamefont {Hallas}},\ }\bibfield  {title} {\bibinfo {title} {Tutorial: a beginner’s guide to interpreting magnetic susceptibility data with the curie-weiss law},\ }\href {https://www.nature.com/articles/s42005-022-00853-y} {\bibfield  {journal} {\bibinfo  {journal} {Communications Physics}\ }\textbf {\bibinfo {volume} {5}},\ \bibinfo {pages} {95} (\bibinfo {year} {2022})}\BibitemShut {NoStop}%
\bibitem [{\citenamefont {Balk}\ \emph {et~al.}(2017)\citenamefont {Balk}, \citenamefont {Kim}, \citenamefont {Pierce}, \citenamefont {Stiles}, \citenamefont {Unguris},\ and\ \citenamefont {Stavis}}]{PhysRevLett.119.077205}%
  \BibitemOpen
  \bibfield  {author} {\bibinfo {author} {\bibfnamefont {A.~L.}\ \bibnamefont {Balk}}, \bibinfo {author} {\bibfnamefont {K.-W.}\ \bibnamefont {Kim}}, \bibinfo {author} {\bibfnamefont {D.~T.}\ \bibnamefont {Pierce}}, \bibinfo {author} {\bibfnamefont {M.~D.}\ \bibnamefont {Stiles}}, \bibinfo {author} {\bibfnamefont {J.}~\bibnamefont {Unguris}},\ and\ \bibinfo {author} {\bibfnamefont {S.~M.}\ \bibnamefont {Stavis}},\ }\bibfield  {title} {\bibinfo {title} {Simultaneous control of the dzyaloshinskii-moriya interaction and magnetic anisotropy in nanomagnetic trilayers},\ }\href {https://doi.org/10.1103/PhysRevLett.119.077205} {\bibfield  {journal} {\bibinfo  {journal} {Phys. Rev. Lett.}\ }\textbf {\bibinfo {volume} {119}},\ \bibinfo {pages} {077205} (\bibinfo {year} {2017})}\BibitemShut {NoStop}%
\bibitem [{\citenamefont {Bouaziz}\ \emph {et~al.}(2024)\citenamefont {Bouaziz}, \citenamefont {Bihlmayer}, \citenamefont {Patrick}, \citenamefont {Staunton},\ and\ \citenamefont {Bl\"ugel}}]{PhysRevB.109.L201108}%
  \BibitemOpen
  \bibfield  {author} {\bibinfo {author} {\bibfnamefont {J.}~\bibnamefont {Bouaziz}}, \bibinfo {author} {\bibfnamefont {G.}~\bibnamefont {Bihlmayer}}, \bibinfo {author} {\bibfnamefont {C.~E.}\ \bibnamefont {Patrick}}, \bibinfo {author} {\bibfnamefont {J.~B.}\ \bibnamefont {Staunton}},\ and\ \bibinfo {author} {\bibfnamefont {S.}~\bibnamefont {Bl\"ugel}},\ }\bibfield  {title} {\bibinfo {title} {Origin of incommensurate magnetic order in the $r\mathrm{AlSi}$ magnetic weyl semimetals $(r=\mathrm{Pr}, \mathrm{Nd}, \mathrm{Sm})$},\ }\href {https://doi.org/10.1103/PhysRevB.109.L201108} {\bibfield  {journal} {\bibinfo  {journal} {Phys. Rev. B}\ }\textbf {\bibinfo {volume} {109}},\ \bibinfo {pages} {L201108} (\bibinfo {year} {2024})}\BibitemShut {NoStop}%
\bibitem [{\citenamefont {Wu}\ \emph {et~al.}(2023)\citenamefont {Wu}, \citenamefont {Xiang},\ and\ \citenamefont {Dai}}]{wu2023tunable}%
  \BibitemOpen
  \bibfield  {author} {\bibinfo {author} {\bibfnamefont {J.}~\bibnamefont {Wu}}, \bibinfo {author} {\bibfnamefont {Y.}~\bibnamefont {Xiang}},\ and\ \bibinfo {author} {\bibfnamefont {X.}~\bibnamefont {Dai}},\ }\bibfield  {title} {\bibinfo {title} {Tunable broadband compact optical isolator based on weyl semimetal},\ }\href {https://www.sciencedirect.com/science/article/pii/S2211379723000839} {\bibfield  {journal} {\bibinfo  {journal} {Results in Physics}\ }\textbf {\bibinfo {volume} {46}},\ \bibinfo {pages} {106290} (\bibinfo {year} {2023})}\BibitemShut {NoStop}%
\bibitem [{\citenamefont {Cheon}\ \emph {et~al.}(2025)\citenamefont {Cheon}, \citenamefont {Kiani}, \citenamefont {Tu}, \citenamefont {Kumar}, \citenamefont {Duong}, \citenamefont {Kim}, \citenamefont {Sam}, \citenamefont {Wang}, \citenamefont {Kushwaha}, \citenamefont {Ng} \emph {et~al.}}]{cheon2025surface}%
  \BibitemOpen
  \bibfield  {author} {\bibinfo {author} {\bibfnamefont {Y.}~\bibnamefont {Cheon}}, \bibinfo {author} {\bibfnamefont {M.~T.}\ \bibnamefont {Kiani}}, \bibinfo {author} {\bibfnamefont {Y.-H.}\ \bibnamefont {Tu}}, \bibinfo {author} {\bibfnamefont {S.}~\bibnamefont {Kumar}}, \bibinfo {author} {\bibfnamefont {N.~K.}\ \bibnamefont {Duong}}, \bibinfo {author} {\bibfnamefont {J.}~\bibnamefont {Kim}}, \bibinfo {author} {\bibfnamefont {Q.~P.}\ \bibnamefont {Sam}}, \bibinfo {author} {\bibfnamefont {H.}~\bibnamefont {Wang}}, \bibinfo {author} {\bibfnamefont {S.~K.}\ \bibnamefont {Kushwaha}}, \bibinfo {author} {\bibfnamefont {N.}~\bibnamefont {Ng}}, \emph {et~al.},\ }\bibfield  {title} {\bibinfo {title} {Surface-dominant transport in weyl semimetal nbas nanowires for next-generation interconnects},\ }\href {https://arxiv.org/pdf/2503.04621} {\bibfield  {journal} {\bibinfo  {journal} {arXiv preprint arXiv:2503.04621}\ } (\bibinfo {year} {2025})}\BibitemShut {NoStop}%
\end{thebibliography}%

\begin{widetext}

\appendix
\renewcommand{\thesection}{\Alph{section}}
\section{WSM Fermi-arc contribution in Cylindrical Coordinate system}
\label{app}

In this appendix, we provide the exact expressions for different RKKY correlations and their approximated expressions in the cylindrical coordinate system.

 We perform the $\omega$ integral in ~Eq.(\ref{RKKY}), and define,
\begin{align*}
I&=\int_{-\infty}^{\epsilon_{F}} d\omega \dfrac{1}{(\omega-\epsilon_{m,k,\xi}+i \eta)(\omega-\epsilon_{n,k',\xi'}+i \eta)}
\\
&=\int_{-\infty}^{\epsilon_{F}} d\omega\dfrac{(\omega-\epsilon_{m,k,\xi})(\omega-\epsilon_{n,k',\xi'})-\eta^{2}}{[(\omega-\epsilon_{m,k,\xi})^{2}+\eta^{2}][(\omega-\epsilon_{n,k',\xi'})^{2}+\eta^{2}]} \\
&~~~~-i \eta \dfrac{(2\omega-\epsilon_{m,k,\xi}-\epsilon_{n,k',\xi'})}{[(\omega-\epsilon_{m,k,\xi})^{2}+\eta^{2}][(\omega-\epsilon_{n,k',\xi'})^{2}+\eta^{2}]}
\\
&=I_{1}-i\ I_{2},
\end{align*}

where $I_{1}$ and $I_{2}$ is given by,

\begin{equation}\label{Integral1}
\begin{aligned}
I_{1}&=\dfrac{\log\Big[\eta^{2}+(\omega-\epsilon_{m,k,\xi})^{2} \Big]-\log\Big[\eta^{2}+(\omega-\epsilon_{n,k',\xi'})^{2} \Big]}{2(\epsilon_{m,k,\xi}-\epsilon_{n,k',\xi'})} \Big|^{\epsilon_{F}}_{-\infty},
\end{aligned}
\end{equation}

\begin{equation}\label{Integral2}
\begin{aligned}
I_{2}&=\dfrac{-\arctan\Big[\dfrac{\epsilon_{m,k,\xi}-\omega}{\eta}\Big]+\arctan\Big[\dfrac{\epsilon_{n,k',\xi'}-\omega}{\eta}\Big]}{\epsilon_{m,k,\xi}-\epsilon_{n,k',\xi'}} \Big|^{\epsilon_{F}}_{-\infty}.
\end{aligned}
\end{equation}

Using the Eq.~\eqref{RKKY}, \eqref{Integral1} and ~\eqref{Integral2}, after putting the expression for the Green's function in Eq.~\eqref{wsmGfunction} we get the following expressions,

\begin{equation}\label{EqApp1}
\begin{aligned}
&J_{zz}=-\frac{J^{2}}{\pi}(\Delta k)^2\Im\sum_{m,k_{\xi},\xi,n,k'_{\xi'},\xi'}[I_{1}-i I_{2}] e^{i[(m-n)\delta\phi+(k-k')\delta z]}\Big[-v_{m,k,\xi}u^{*}_{m,k,\xi}v^{*}_{n,k',\xi'}u_{n,k',\xi'}-u_{m,k,\xi}v^{*}_{m,k,\xi}u^{*}_{n,k',\xi'}v_{n,k',\xi'}
\\
&+|v_{m,k,\xi}|^{2}|v_{n,k',\xi'}|^{2}+|u_{m,k,\xi}|^{2}|u_{n,k',\xi'}|^{2}\Big],
\end{aligned}
\end{equation}
\begin{equation}
\begin{aligned}
&J_{rz}=-\frac{J^{2}}{\pi}(\Delta k)^2\Im\sum_{m,k_{\xi},\xi,n,k'_{\xi'},\xi'}[I_{1}-i I_{2}] e^{i[(m-n)\delta\phi+(k-k')\delta z]}\Big[v_{m,k,\xi}u^{*}_{m,k,\xi}|u_{n,k',\xi'}|^{2}-|v_{m,k,\xi}|^{2}v_{n,k',\xi'}u^{*}_{n,k',\xi'}+|u_{m,k,\xi}|^{2}
\\
&u_{n,k',\xi'}v^{*}_{n,k',\xi'}-u_{m,k,\xi}v^{*}_{m,k,\xi}|v_{n,k',\xi'}|^{2}\Big],
\end{aligned}
\end{equation}
\begin{equation}
\begin{aligned}
&J_{zr}=-\frac{J^{2}}{\pi}(\Delta k)^2\Im\sum_{m,k_{\xi},\xi,n,k'_{\xi'},\xi'}[I_{1}-i I_{2}] e^{i[(m-n)\delta\phi+(k-k')\delta z]}\Big[u_{m,k,\xi}v^{*}_{m,k,\xi}|u_{n,k',\xi'}|^{2}-|v_{n,k',\xi'}|^{2}v_{m,k,\xi}u^{*}_{m,k,\xi}+|u_{m,k,\xi}|^{2}
\\
&v_{n,k',\xi'}u^{*}_{n,k',\xi'}-u_{n,k',\xi'}v^{*}_{n,k',\xi'}|v_{m,k,\xi}|^{2}\Big],
\end{aligned}
\end{equation}

\begin{equation}
\begin{aligned}
&J_{rr}=-\frac{J^{2}}{\pi}(\Delta k)^2\Im\sum_{m,k_{\xi},\xi,n,k'_{\xi'},\xi'}[I_{1}-i I_{2}] e^{i[(m-n)\delta\phi+(k-k')\delta z]}\Big[v_{m,k,\xi}u^{*}_{m,k,\xi}v_{n,k',\xi'}u^{*}_{n,k',\xi'}+u_{m,k,\xi}v^{*}_{m,k,\xi}u_{n,k',\xi'}v^{*}_{n,k',\xi'}
\\
&+|v_{m,k,\xi}|^{2}|u_{n,k',\xi'}|^{2}+|u_{m,k,\xi}|^{2}|v_{n,k',\xi'}|^{2}\Big],
\end{aligned}
\end{equation}
\begin{equation}
\begin{aligned}
&J_{\phi\phi}=-\frac{J^{2}}{\pi}(\Delta k)^2\Im\sum_{m,k_{\xi},\xi,n,k'_{\xi'},\xi'}[I_{1}-i I_{2}] e^{i[(m-n)\delta\phi+(k-k')\delta z]}\Big[-v_{m,k,\xi}u^{*}_{m,k,\xi}v_{n,k',\xi'}u^{*}_{n,k',\xi'}-u_{m,k,\xi}v^{*}_{m,k,\xi}u_{n,k',\xi'}v^{*}_{n,k',\xi'}
\\
&+|v_{m,k,\xi}|^{2}|u_{n,k',\xi'}|^{2}+|u_{m,k,\xi}|^{2}|v_{n,k',\xi'}|^{2}\Big],
\end{aligned}
\end{equation}
\begin{equation}
\begin{aligned}
&J_{r\phi}=-\frac{J^{2}}{\pi}(\Delta k)^2\Im\sum_{m,k_{\xi},\xi,n,k'_{\xi'},\xi'}[I_{1}-i I_{2}] e^{i[(m-n)\delta\phi+(k-k')\delta z]}\Big[-iv_{m,k,\xi}u^{*}_{m,k,\xi}v_{n,k',\xi'}u^{*}_{n,k',\xi'}+i u_{m,k,\xi}v^{*}_{m,k,\xi}u_{n,k',\xi'}v^{*}_{n,k',\xi'}\\
&+i|v_{m,k,\xi}|^{2}
|u_{n,k',\xi'}|^{2}-i|u_{m,k,\xi}|^{2}|v_{n,k',\xi'}|^{2}\Big],
\end{aligned}
\end{equation}
\begin{equation}
\begin{aligned}
&J_{\phi r}=-\frac{J^{2}}{\pi}(\Delta k)^2 \Im\sum_{m,k_{\xi},\xi,n,k'_{\xi'},\xi'}[I_{1}-i I_{2}] e^{i[(m-n)\delta\phi+(k-k')\delta z]}\Big[-iv_{m,k,\xi}u^{*}_{m,k,\xi}v_{n,k',\xi'}u^{*}_{n,k',\xi'}+i u_{m,k,\xi}v^{*}_{m,k,\xi}u_{n,k',\xi'}v^{*}_{n,k',\xi'}\\
&-i|v_{m,k,\xi}|^{2}
|u_{n,k',\xi'}|^{2}+i|u_{m,k,\xi}|^{2}|v_{n,k',\xi'}|^{2}\Big],
\end{aligned}
\end{equation}
\begin{equation}
\begin{aligned}
&J_{z\phi}=-\frac{J^{2}}{\pi}(\Delta k)^2\Im\sum_{m,k_{\xi},\xi,n,k'_{\xi'},\xi'}[I_{1}-i I_{2}] e^{i[(m-n)\delta\phi+(k-k')\delta z]}\Big[i u_{m,k,\xi}v^{*}_{m,k,\xi}|u_{n,k',\xi'}|^{2}+i v_{m,k,\xi}u^{*}_{m,k,\xi}|v_{n,k',\xi'}|^{2}-i|u_{m,k,\xi}|^{2}
\\
&v_{n,k',\xi'}u^{*}_{n,k',\xi'}-i|v_{m,k,\xi}|^{2}u_{n,k',\xi'}v^{*}_{n,k',\xi'}\Big],
\end{aligned}
\end{equation}

\begin{equation}\label{EqApp9}
\begin{aligned}
&J_{\phi z}=-\frac{J^{2}}{\pi}(\Delta k)^2\Im\sum_{m,k_{\xi},\xi,n,k'_{\xi'},\xi'}[I_{1}-i I_{2}] e^{i[(m-n)\delta\phi+(k-k')\delta z]}\Big[-i v_{m,k,\xi}u^{*}_{m,k,\xi}|u_{n,k',\xi'}|^{2}-i u_{m,k,\xi}v^{*}_{m,k,\xi}|v_{n,k',\xi'}|^{2}
\\
&+i|v_{m,k,\xi}|^{2}v_{n,k',\xi'}u^{*}_{n,k',\xi'}+i|u_{m,k,\xi}|^{2}u_{n,k',\xi'}v^{*}_{n,k',\xi'}\Big].
\end{aligned}
\end{equation}

\begin{figure}
\centering\includegraphics[width=0.8\linewidth]{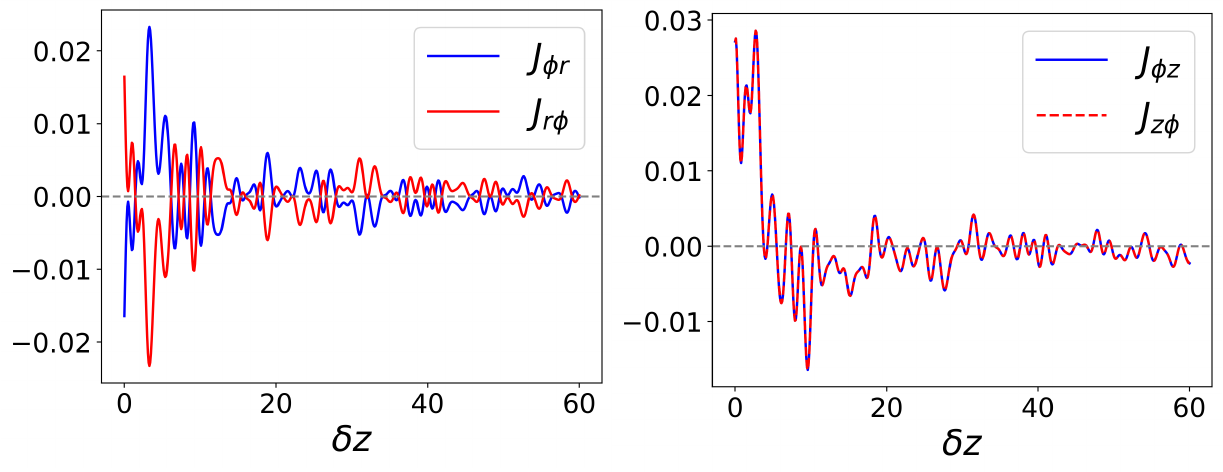}
\caption{The correlation $J_{\phi z},J_{z \phi}, J_{r \phi}, J_{\phi r}$ are plotted as a function of $\delta z$, with $\epsilon_{F}=0.44$, all the other parameter is same as in Fig~(\ref{wsmcorr1}). These correlations become zero analytically when we take only the contributions coming from the Fermi-arc surface states.}
\label{fig:appendix-figure1}
\end{figure}

\subsubsection*{Mathematical functions}
The Bessel function of the first kind is defined as,

\begin{equation}
J_{m}(x)=\sum_{l=0}^{\infty}\dfrac{(-1)^l}{l! \Gamma(m+l+1)}\Big(\dfrac{x}{2}\Big)^{2l+m},
\end{equation}

where, $\Gamma(z)$ is the gamma function. The Bessel function of the second kind is defined as,

\begin{equation}
Y_{m}(x)=\dfrac{J_{m}(x)\cos(m \pi)-J_{-m}(x)}{\sin(m \pi)},
\end{equation}

and the Hankel function of the first kind if given by,

\begin{equation}
H_{m}^{(1)}(x)=J_{m}(x)+iY_{m}(x).
\end{equation}

\begin{comment}
\begin{equation*}
A_{m,n,\xi,\xi'}(k,k',\omega)=\dfrac{\theta(k-k_{\xi,1}) \theta(k_{\xi,2}-k)\theta(k'-k'_{\xi',1}) \theta(k'_{\xi',2}-k')}{\big(\omega-(m+ \dfrac{1}{2})\ \xi \dfrac{v_{F}}{R}+i\eta\big)\big(\omega-(n+\dfrac{1}{2})\ \xi' \dfrac{v_{F}}{R}+i\eta\big)}.
\end{equation*}
\end{comment}

\end{widetext} 

\end{document}